\documentclass[pre,aps,superscriptaddress,twocolumn]{revtex4-1}   	

\usepackage{amsmath}
\usepackage{graphicx}
\usepackage{subfigure}
\usepackage{soul}
\usepackage{bbold}
\usepackage{color}
\usepackage{booktabs}
\usepackage{array} 
\usepackage{mathtools}
\usepackage[export]{adjustbox}
\graphicspath{{figures/}}

\usepackage{dcolumn}
\usepackage{makecell}

\newcommand{\beq}{\begin{equation}}
\newcommand{\eeq}{\end{equation}}
\newcommand{\vv}[1]{\left(\begin{array}{c}#1\end{array}\right)}
\newcommand{\p}{\partial}

\begin{document}

\title{Geometrically-induced localization of flexural waves on thin warped physical membranes}

\author{Jonathan Kernes}
\affiliation{Department of Physics and Astronomy, UCLA, Los Angeles California 90095-1596, USA}
\author{Alex J. Levine}
\affiliation{Department of Physics and Astronomy, UCLA, Los Angeles California 90095-1596, USA}
\affiliation{Department of Chemistry and Biochemistry, UCLA, Los Angeles California 90095-1596, USA}
\affiliation{Department of Computational Medicine, UCLA, Los Angeles California 90095-1596, USA}


\begin{abstract}
We consider the propagation of flexural waves across a nearly flat, thin membrane, whose stress-free state is curved. The stress-free configuration is specified by a quenched height field, whose Fourier components are drawn from a Gaussian distribution with power law variance. Gaussian curvature couples the in-plane stretching to out-of-plane bending. Integrating out the faster stretching modes yields a wave equation for undulations in the presence of an effective random potential, determined purely by geometry. We show that at long times/lengths, the undulation intensity obeys a diffusion equation. The diffusion coefficient is found to be frequency dependent and sensitive to the quenched height field distribution. Finally, we consider the effect of coherent backscattering corrections, yielding a weak localization correction that decreases the diffusion coefficient proportional to the logarithm of the system size, and induces a localization transition at large amplitude of the quenched height field. The localization transition is confirmed via a self-consistent extension to the strong disorder regime.

\end{abstract}
\maketitle

\section{Introduction}

Thin, elastic shells, whose lateral size is much larger than their thickness, arise in a diverse array of contexts and across systems spanning several orders of magnitude in length~\cite{kearey2009global,wang2005flexural,bower1999deformation}. Though the initial theory was formulated more than a century ago, shells that retain curvature in the stress-free state exhibit complex solutions due to geometrically induced nonlinearity, which has continued to inspire research~\cite{vaziri2008localized,lazarus2012geometry,bende2015geometrically}. The mechanical properties of shells with curved stress-free states is vital to the functionality of a number of biological applications, including red blood cells~\cite{waugh1979thermoelasticity,park2010measurement,park2011measurement}, viral capsids~\cite{lidmar2003virus,ivanovska2004bacteriophage,kuriabova2008nanorheology,michel2006nanoindentation,klug2006failure}, and plant morphogenesis~\cite{dervaux2008morphogenesis}. Though these latter examples lack well defined elastic continua, they have nevertheless been successfully studied using thin shell theory.

The mechanics of thin shells are determined by geometry, and shells of different curvature in the stress-free state can exhibit wildly different characteristics~\cite{radzihovsky1991statistical,morse1992curvature,radzihovsky1992crumpled,le1993flat}. Due to Gauss' {\it theorema egregium}, Gaussian curvature couples the typically soft bending undulations to the much stiffer stretching deformations. As a result, areas of positive Gauss curvature suppress bending undulations~\cite{vaziri2008localized}, which can lead to spatially heterogenous pockets of large undulations separated by boundary curves of zero normal curvature, as has been observed in the fluctuations of red blood cells~\cite{evans2017geometric}. As undulations typically represent the softest elastic modes, these can have serious consequences on the ability for shells to equilibrate, which has applications for the stability of cell membranes with actively driven pumps.

An examination of undulatory waves in the geometric optics limits shows that Gauss curvature scatters undulatory waves, and can even lead to total internal reflection at boundaries where the sign changes~\cite{evans2013reflection}. This suggests that for thin shells with random stress-free curvature, energy transport could be severely slowed, if not localized, purely as a consequence of geometry.

In this manuscript, we present an analysis of the propagation of undulatory waves through randomly curved thin shells that, in contrast to the large literature of wave propagation in random media~\cite{ishimaru2017electromagnetic}, is driven entirely by geometry. Wave propagation in random media has been extensively studied~\cite{akkermans2007mesoscopic,rammer2018quantum,van1999multiple,sheng2006introduction,chakravarty1986weak} across a broad array of subjects~\cite{campillo2003long,korn1993seismic,bayer1993weak}, which we use as a guide.

The outline of this paper is as follows. First, a general consequence of random scattering is to shift energy propagation from ballistic to diffusive, which we quantitatively compute in the weak scattering limit. This is done via a hydrodynamic derivation, looking at the long length/time limits of energy transport~\cite{sheng2006introduction,vollhardt1992self,barabanenkov1991asymptotic}. In Appendix~\ref{app: diffuson}, we present an alternative diagrammatic derivation in terms of the well-known ladder diagram approximation~\cite{akkermans2007mesoscopic,rammer2018quantum,van1999multiple,sheng2006introduction,chakravarty1986weak}. Time reversal invariant systems possess an additional contribution in the hydrodynamic limit, {\it weak localization}~\cite{akkermans1985weak,wolf1985weak,van1985observation}, that serves to inhibit diffusion. After computing the weak localization correction, we consider the behavior of waves in the strong scattering regime. At strong scattering, waves can undergo a localization transition~\cite{anderson1958absence,thouless1974electrons}, whereby they are spatially localized even in the absence of energy barriers. We extend our weak localization result to the strong scattering limit via a self-consistent condition~\cite{vollhardt1992self}, where we find that undulatory waves exhibit the same exponentially large localization length endemic to other two dimensional systems, such as electrons in a random potential~\cite{sheng2006introduction}.

Finally, we attempt to summarize our results in the limits of waves propagating through large/small disorder where the weighty expressions simplify. 

\section{Generalized Donnel-Mushtari-Vlasov (DMV) linearized shallow shell theory}

We define membranes as a particular class of thin shells. Membranes are elastic media with two internal dimensions describing in-plane stretching deformations, and $d_c$ surface normals describing the direction of bending undulations, embedded in a $d_c+2$ dimensional space. 
Throughout this manuscript, we employ the convention that Greek indices correspond to the $d_c$ normal directions, and Latin indices to the two internal dimensions. Bold-face letters refer to vectors in the $(d_c+2)$ dimensional embedding space. 

The purpose of this generalization to arbitrary embedding dimension is to later allow us to use the self-consistent screening approximation (SCSA) to partially resum perturbation series encountered upon disorder averaging (see Appendix~\ref{app: self-energy})~\cite{kovsmrlj2013mechanical,kovsmrlj2014thermal}. Ultimately, we are interested in the physically realizable case of $d_c=1$, which we hereafter refer to as physical membranes. 

To isolate the role of geometry, we focus our analysis on warped membranes~\cite{kovsmrlj2013mechanical,kovsmrlj2014thermal} (for behavior of these membranes under thermal fluctuations see Refs~\cite{radzihovsky1991statistical,radzihovsky1992crumpled,morse1992curvature,le1993flat}); these are nearly flat membranes of internal volume $L^2$, with stress-free local height configuration that can be given in the Monge representation~\cite{nelson2004statistical} by a quenched, random background height field $h_\beta({\bf x})$. Specifically, the stress-free membrane is described by the vector

\begin{figure}
\includegraphics[scale=0.8]{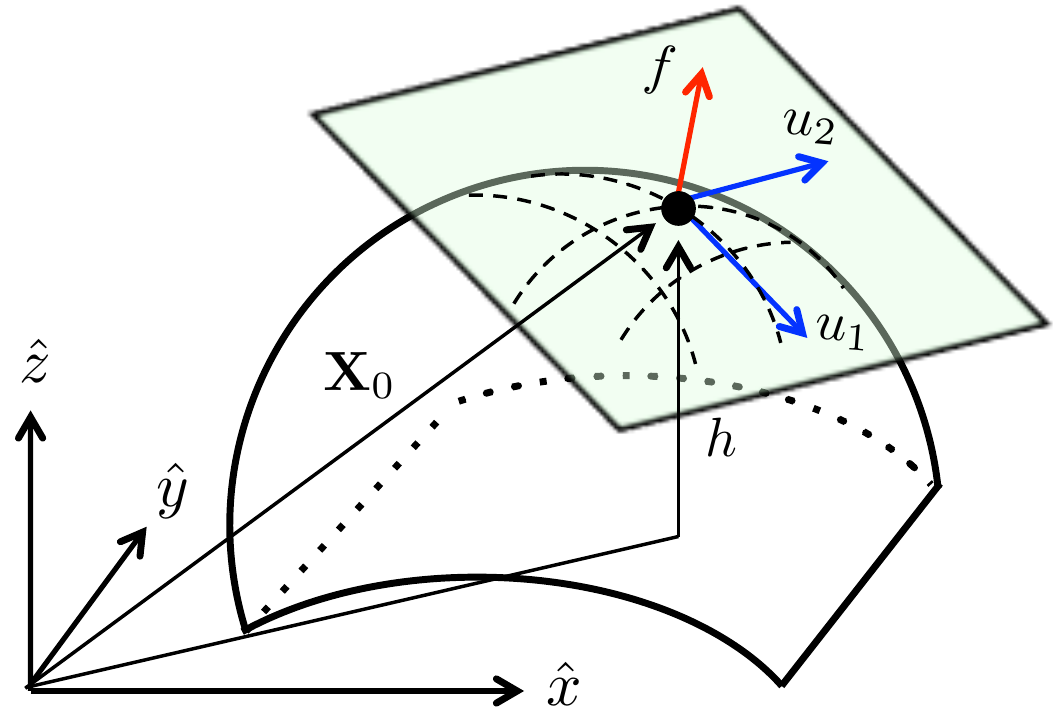}
\caption{(Color online) parametrization of a physical membrane middle surface ${\bf X}_0$ in the Monge representation, with small deformations $f,\,u_1,$ and $u_2$ given in normal coordinates. Displacements $u_1,u_2$ (in blue) are along the local surface tangent of the curved background surface, while displacements $f$ (in red) are in the direction of the local surface normal. Misalignment of the local surface normal with the global $\hat z$ direction is responsible for additional curvature terms in the strain (see Eq.~\ref{linear strain}).}
\label{fig: schematic3}
\end{figure}

\beq
\label{Coordinate vector}
{\bf{X}}_0(x_a) = x_a {\bf{t}}_a + h_\beta{\bf{n}}_\beta,
\eeq
where ${\bf t}_a$ represent the local surface tangents, and $\bf{n}_\beta$ the local surface normals. The total displacement field after small deformation is denoted ${\bf X}(x_a)$. The quenched height field is a random variable, whose Fourier coefficients
\beq
\label{eq: Fourier h}
h({ \bf p }) = \frac{1}{L^2}\int d^2 x \, h({\bf x})e^{-i {\bf p} \cdot {\bf x}},
\eeq
are sampled from a Gaussian distribution with zero mean, and variance
\beq
\label{disorder correlation}
\langle h_\alpha( {\bf p} ) h_\beta ( {\bf p}') \rangle = \frac{ \gamma\delta_{\alpha \beta}\delta_{{\bf p}, -{\bf p'}}}{L^2  {\bf p}^{2 d_H}}.
\eeq
The parameter $\gamma$ specifies the amplitude of the quenched height field and determines the strength of disorder. We focus on the experimentally relevant cases $d_H =0, 2,4$, which may be realized in biological contexts by rapid polymerization of fluctuating lipid bilayers~\cite{Larche1986,Safinya1986,kovsmrlj2014thermal}.

To quadratic order, the change in potential energy about a particular configuration ${\bf X}(x_a)$ is~\cite{kovsmrlj2013mechanical}
\beq
\begin{split}
\label{eq: elastic energy}
\mathcal{U} = \frac{1}{2}\int d^2 x \left\{ 2\mu (U_{ab})^2 + \lambda (U_{cc})^2 + \kappa (K_{cc})^2\right\}.
\end{split}
\eeq
The tensors $U_{ab}$ and $K_{ab}$ represent the variation in the metric tensor and generalized bending tensor respectively, from their background configurations;

\beq
\label{strain tensors}
U_{ab} = \p_a { \bf X} \cdot \p_b { \bf X} - \delta_{ab}, \;\;\; {\bf K}_{ab} = \p_ a \p_b { \bf X} -{\bf d}_{ab},
\eeq
and we have defined the quenched background curvature tensor 
\beq
\label{eq: background d tensor}
{\bf d}_{ab} = (\p_a \p_b h_\beta ){\bf n}_\beta.
\eeq
Ignoring small deformations of $\mathcal{O}((\p_a h_\beta)^2)$, the background metric tensor is flat ($\delta_{ab}$). The background curvature tensor (second fundamental form~\cite{niordson2012shell,frankel2011geometry}) then encodes all effects of the disordered geometry.

We decompose the deformation vector ${\bf X}={\bf X}_0+\delta {\bf X}$ into a background configuration ${\bf X}_0$ plus small deformation $\delta {\bf X}$. The latter is parametrized in normal coordinates~\cite{nelson2004statistical}, separating local strain into in-hyperplane displacements, $ u_a(x_a)$, and out-of-hyperplane displacements, $f_\beta(x_a)$, via

\beq
\label{Displacement vector}
\delta {\bf X} = u_a {\bf  t}_a + f_\beta {\bf n}_\beta.
\eeq

The equations of motion derived from Eq.~\ref{eq: elastic energy} are formidable. We work in the limit where variations in curvature are slow on the scale of characteristic deformation wavelength. This is summarized as the condition $w \ll \lambda \ll |R|$, for $w$ the membrane thickness, $\lambda$ the characteristic deformation wavelength, and $R$ the average radius of curvature. Utilizing the definition of the tangent vector, ${\bf t}_a= \p_a{ \bf X}$, this leads to the linearized strain fields

\beq
\label{linear strain}
U_{ab}= \frac{1}{2}(\p_a u_b+\p_b u_a) -  f_\alpha \p_a\p_b h_\alpha, \;\; { \bf K}_{ab} = (\p_a \p_b f_\alpha ){\bf n}_\alpha.
\eeq
This is the lowest order coupling of bending undulations to Gauss curvature. Deformations $f_\beta$ are considered small compared to $h_\beta$, and hence nonlinear terms may be neglected. In terms of the fields $u_a,f_\beta$, the elastic energy now reads:

\beq
\label{eq: elastic energy full}
\begin{split}
\mathcal{U} &=\frac{1}{2}\int d^2 x \left\{\kappa (\nabla^2 f_\beta)^2 +  u_a \nabla^2 \left((\lambda+2\mu) \hat P_{ab}^L + \mu \hat P_{ab}^T \right) u_b \right. \\
&+2\mu \text{Tr}\left(\left( \hat P^T_{ab} f_\alpha \p_b \p_c h_\alpha\right) ^2\right) + \lambda  (\hat P^T_{ab} f_\alpha \p_a \p_b h_\alpha)^2  \\
&\left.+ 2\lambda \p_a u_a (\hat P^T_{bc} f_\alpha \p_b \p_c h_\alpha)\right \} ,
\end{split}
\eeq
where we have defined the projection operators 
\beq
\hat P_{ij}^L = \nabla^{-2} \p_i \p_j, \qquad \hat P^T_{ij} = \delta_{ij} - \hat P_{ij}^L.
\eeq

The first line of Eq.~\ref{eq: elastic energy full} is the elastic energy of a flat plate. In-hyperplane deformations split into longitudinal and transverse components. The second line penalizes bending in regions of positive Gauss curvature. The third line is the linear coupling between bending and stretching, which can allow undulations to {\it tunnel} through regions of positive Gauss curvature.
Only the longitudinal component of $u_a(x_a)$ couples. 

To find dynamical solutions, we must form an action by including a kinetic energy density. Since in-hyperplane displacements relax much faster than bending undulations, we approximate that the $u_a$ fields respond instantaneously to deformation. We therefore only include an undulatory kinetic energy density $(\sigma/2)(\p_t f_\beta)^2$, for $\sigma$ the surface mass density. Furthermore, since $u_a$ is simply a constraint field, we may eliminate it by setting it equal to its equation of motion, yielding an effective action describing the dynamics of undulatory waves subject to a long range potential. As the membrane length $L$ is considered large, its bulk properties do not depend on boundary condition, which we will assume to be periodic. We switch to and from Fourier space via
\begin{subequations}
\beq
\label{eq: Fourier sum}
f({\bf x}) = \sum_{\bf p} f_{\bf p} e^{i {\bf p} \cdot {\bf x}},
\eeq
\beq
f_{\bf p} = \frac{1}{L^d} \int d^d x f({\bf x}) e^{-i {\bf p} \cdot {\bf x}}.
\eeq
\end{subequations}
Here, and for the remainder of the manuscript, bold face refers to vectors in the internal two-dimensional space. The summations run over $p_1,p_2 = 2 \pi n/L$, for $n \in \mathbb{Z}$.  In Fourier space we find the effective undulation action~\cite{radzihovsky1991statistical,kovsmrlj2014thermal}

\begin{align}
\label{eq: S}
S &=S_0 + S_{\text{int}}, \\ 
\label{eq: S0}
S_0 &= \frac{L^2}{2}\sum_{{\bf q}}  \left(\kappa q^4 - \sigma \omega^2\right) f_{{\bf q}} f_{{\bf -q}},  \\
\label{eq: Sint}
S_{\text{int}}&= \sum_{ \substack{{\bf q}_2+{\bf q}_4=-{\bf q} \\ {\bf q}_1 + {\bf q}_3 ={\bf q}} } 
\frac{L^2}{d_c}{\bf q}_1^i{\bf q}_1^jR_{ij,kl}({\bf q}){\bf q}_2^k{\bf q}_2^l h^\alpha_{{\bf q}_1}h^\beta_{{\bf q}_2}f^\alpha_{{\bf q}_3}f^\beta_{{\bf q}_4},
\end{align}
where we have defined the interaction tensor
\beq
\hat R_{ij,kl}=  \frac{\tilde \mu}{2} \left( \hat P_{ik}^T \hat P_{jl}^T + \hat P_{il}^T \hat P_{jk}^T + \frac{2 \tilde \lambda}{\tilde \lambda+2 \tilde \mu} \hat P_{ij}^T \hat P_{kl}^T\right).
\eeq
This form is valid for arbitrary internal dimension. For physical membranes with two internal dimensions, the indices are irrelevant and we can write the simpler form
\beq
\label{eq: R def}
R_{ij,kl}(q)= \frac{Y}{2d_c} P_{ik}^T(q) P_{jl}^T(q),
\eeq
where $Y$ is the two dimensional Young's modulus
\beq
\label{eq: Y}
Y =\frac{4 \mu(\lambda + \mu)}{\lambda + 2\mu}.
\eeq
 
The equations of motion are typically written including the displacements $u_a$, and are derived by variation of the elastic energy in Eq.~\ref{eq: elastic energy full}. These, in turn, are further reduced by eliminating $u_a$ in terms of a scalar Airy stress function $\chi$, defined by its relation to the stress tensor $\delta \mathcal{U}/\delta U_{ab} = \epsilon_{ac}\epsilon_{bd} \p_c \p_d \chi$. In terms of $\chi$ and $f$, we find the Donnell-Mushtari-Vlasov (DMV) linearized shallow shell equations~\cite{niordson2012shell,evans2013reflection,vaziri2008localized}

\begin{eqnarray}
\label{eq: DMV1}
\frac{1}{Y} \nabla^4 \chi + \mathcal{\hat L} f &=&0  \\
\label{eq: DMV2}
\kappa \nabla^4 f - \sigma \frac{\partial^2 f}{\partial t^2} - \mathcal{\hat L} \chi &=&0.
\end{eqnarray}
The operator
\beq
\label{eq: L def}
\mathcal{\hat L} = \epsilon_{ac}\epsilon_{bd} (\p_a \p_b h)\p_c \p_d,
\eeq
contains the quenched random height field, and encodes undulatory scattering due to curvature. Though for elastic continua $Y$ represents a Young's modulus and $\kappa$ a bending rigidity, these may alternatively be viewed as phenomenological parameters governing the strength of stretching and bending respectively when the membrane is not derivable from elastic continua. Eqs.~\ref{eq: DMV1},~\ref{eq: DMV2} represent the main equations of motion for physical membranes in linearized shallow shell theory.

By formally integrating out the fields $\chi$, we find the $f$ field experiences an effective potential $\hat V =\mathcal{ \hat L}\nabla^{-4} \mathcal{ \hat L}$. Via the Gauss-Bonnet theorem~\cite{nelson2004statistical}, the Riemann curvature is equal to twice the Gauss curvature. Since the metric is flat, the total Gauss curvature is then given by the determinant of the total bending tensor ${\bf d}_{ab}+{\bf K}_{ab}$ (these are $d_c+2$ dimensional vectors). To linear order in $f$, $\delta R({\bf x},t)=\mathcal{\hat L} f$ represents the local change in Gauss curvature. Geometry can be viewed as inducing a two-body $\nabla^{-4}$ potential acting between different regions of Gauss curvature, attractive/repulsive between opposite/same sign, as a response to the membrane trying to flatten~\cite{nelson1987fluctuations}.

\section{Signatures of localization}
\label{sec: signatures}

We now illustrate several quantities that measure the degree of localization and diffusion for undulatory waves. Conserved quantities are particularly useful, as they retain information about an initial injected disturbance at both long times and after disorder averaging, and have the potential for diffusive dynamics. Their propagation through the membrane (or lack thereof) acts as a marker for the spatial localization of waves. 

We give two examples. First, the equations of motion are time reversal invariant, indicating total energy is conserved. Transport of energy across the membrane gives information on heat transfer and the thermal conductivity of the system, both of which are of interest in mechanics of cellular membranes~\cite{evans2017geometric}. Second, the Gauss-Bonnet theorem stipulates that the integral of the Riemann curvature over the membrane is a fixed, topological value, identical over all instances of disorder. For physical membranes, the Gauss curvature is proportional to the Riemann curvature and hence can play the role of a conserved quantity.

An additional quantity to study is the kinetic energy, which for monochromatic waves is proportional to the disorder averaged local intensity $\langle |f_\beta({\bf x})|^2 \rangle$. We will find that undulation intensity obeys diffusive dynamics, and is the primary cause of diffusive energy/curvature transport. We thus focus our efforts on intensity, which is fortunate, as working with the total energy is analytically challenging.

In order to describe wave propagation, we translate the problem into the language of Green's functions~\cite{john1983localization,kirkpatrick1985localization}. We consider a physical membrane ($d_c=1$) for ease of notation, with generalization to higher dimensionality straightforward. Without loss of generality, the equation of motion may be written as $(\p_t^2 - \hat A) f({\bf x}, t)=j({\bf x},t)$, for some operator $\hat A$ derived from minimization of the action in Eq.~\ref{eq: S}, and arbitrary undulation source $j({\bf x},t)$. Associated with this is a Green's function, $G(x,x';t,t')$, that satisfies the equation $(\p_t^2 - \hat A)G(x,x';t,t')  = \delta(x-x')\delta(t-t')$. 

We are interested in the source-free situation where at times $t<0$, the membrane is in mechanical equilibrium ($f({\bf x},0)=0$), then at $t=0$ it is plucked ($\p_t f({\bf x},0) =1$) with unit velocity at the origin, thereby injecting energy into the system localized at the origin. This is accomplished in the Green's function formalism by imposing initial conditions~\cite{kirkpatrick1985localization}
\begin{eqnarray}
\p_tG^+({\bf x},{\bf x}';0) &=& \delta({\bf x}-{\bf x}'), \\
G^+({\bf x},{\bf x}';t<0) &=& 0.
\end{eqnarray}
Provided the initial pluck is truly $\delta$-like, the main result is that we can everywhere replace the time-dependent amplitude by a Green's function via
\beq
\label{amplitude to Greens function}
f({\bf x},t) = G^+({\bf x},0;t).
\eeq
We now define the disorder averaged quantity $\phi({\bf x},t)$, which represents the intensity of undulations across the membrane in response to an initial pluck at the origin:
\beq
\label{eq: phi real}
\phi(x,t) = \langle G^+({\bf x},0;t) G^-(0,{\bf x};t) \rangle.
\eeq
Here and throughout, angular brackets refer to disorder averaging over the quenched random height field. Our principle object of study is the four-point function
\beq
\label{eq: four point}
\phi_{pp'}({\bf q},\Omega) \equiv \langle G_{\omega_+}^+({\bf p}_+,{\bf p}'_+)G_{\omega_-}^-({\bf p}_-,{\bf p}_-')\rangle,
\eeq
which gives the Fourier components of $\phi({\bf x},t)$ upon summation over $p,p'$ and integration of $\omega$. We have introduced the shorthand $p_\pm \equiv p \pm q/2$ and $\omega_\pm \equiv \omega \pm \Omega/2$. We refer to $\omega$ as the ${\it internal}$ frequency, and $\Omega$ as the {\it external} frequency. The four-point function is the product of an advanced and retarded propagator, and has the necessary phase cancellation to propagate across multiple scattering processes. In terms of the four-point function, the disorder averaged kinetic energy density is
\beq
T({\bf x},\Omega) = \frac{1}{2}\int \frac{d\omega}{2\pi} (\omega^2 +\Omega^2/4) \sum_{pp'q}e^{i {\bf q} \cdot {\bf x}} \phi_{pp'}({\bf q},\Omega).
\eeq

We measure the degree of localization by the normalized spatial variance of kinetic energy in the long-time limit. For diffusive waves, the variance grows linearly in time, while for propagating waves it grows quadratically. This leads us to define the kinetic energy diffusion coefficient~\cite{john1983localization}
\beq
\label{eq: DE time}
D_E = \lim_{t \to \infty}\frac{1}{t}\frac{\int x^2 T(x,t) dx}{\int T(x,t) dx}.
\eeq
We enact the long time limit by means of the identity~\cite{mckane1981localization}
\beq
\lim_{t\to \infty} f(t) =\lim_{\eta \to0} \eta \int f(t) e^{-\eta t} dt.
\eeq 
Upon Fourier transform, we obtain
\beq
\label{eq: DE freq}
D_{E}= \lim_{\eta \to0}\frac{-2\eta^2 \int \frac{d\omega}{2\pi} \omega^2 \nabla_{\bf q}^2|_{q=0}  \sum_{kk'}\phi_{kk'}(q,-2i\eta)}
{\eta \int \frac{d\omega}{2\pi} \omega^2 \sum_{kk'} \phi_{kk'}(q,-2i\eta)}
\eeq

In general, the frequency integrations will diverge as a consequence of the $\delta$-function singularity of the initial spike. This can be regulated by replacing the $\delta$-function with a Gaussian impulse at $t=0$ of width $\Delta t$. The integrations are regulated by simply adjusting the measure $d\omega \to d\omega e^{-\frac{1}{2}\omega^2 \Delta t}$. If the disorder is short-ranged, corresponding to $d_H=0$, high frequencies are cutoff as a consequence of finite lattice spacing, and this may be the more relevant cutoff for short impulses. 

We can further simplify $D_E$ by anticipating that at small frequency $\Omega$, the four-point function is diffusive. Indeed, borrowing the later result in Eq.~\ref{eq: phi},
\beq
\sum_{pp'} \phi_{pp'}=\frac{\pi \nu/(2 \sigma^2 \omega^2 L^2)}{-2 \eta + D(\omega) q^2}.
\eeq

we obtain the much simpler form
\beq
\label{eq: DE final}
D_{E} = 4 \frac{ \int d \omega \, \nu(\omega) D(\omega)}{\int d\omega \, \nu(\omega)},
\eeq
where the frequency dependence of $\nu$ has been made explicit. The function $\nu$ represents the density of states, and suggests the quantity $D(\omega)$ is a frequency dependent intensity diffusion coefficient. 

In fact, $D(\omega)$ corresponds precisely to the diffusion coefficient of intensity in response to a harmonic, transverse load applied at the origin. To show this, we note that the four-point function corresponding to harmonic loading at the origin is equivalent to that of Eq.~\ref{eq: four point}, so long as we take $\Omega \to -2i \eta$ and pin the internal frequency to $\omega$ as opposed to integrating. The variance in intensity is then given by Eq.~\ref{eq: DE freq} if we apply the same conditions, which after simplification yields the frequency dependent diffusivity

The calculation of $D(\omega)$ is a point of contact with numerical simulations that are solved in the frequency domain. Specifically, given knowledge of the solution $f({\bf x}, \omega)$ in response to harmonic loading, one identifies
\beq
D(\omega) = -\frac{1}{2}\lim_{\eta \to 0}\eta \frac{ \int d^2 x x^2 |f({\bf x}, \omega - i \eta)|^2}{\int d^2 x |f({\bf x}, \omega - i \eta)|^2}.
\eeq

\section{Hydrodynamics}
\label{sec: hydrodynamics}

The four-point function defined in Eq.~\ref{eq: four point} is the fundamental quantity governing diffusion/localization. In this section we present a derivation of its diffusive behavior at long length/time scales.

The four point function is a disorder averaged object, which may in general be decomposed into a sum of independently averaged Green's functions, plus all connected averages. Implementing this procedure leads to the Bethe-Salpeter (BS) equation~\cite{vollhardt1992self}
\begin{eqnarray}
\label{eq: BS}
\phi_{pp'}(q,\Omega) &=& \langle G_{p_+}^+(\omega_+) \rangle \langle G_{p_-}^-(\omega_-) \rangle \times  \\ \nonumber
 && \left ( \delta_{pp'}+ \sum_k U_{pk}(q,\Omega) \phi_{kp'}(q,\Omega)  \right).
\end{eqnarray}
The lack of a summation over frequencies is a consequence of quenched disorder. The frequency dependence can be grouped into a single three-vector $({\bf p},\omega)$. When not explicitly written, the three-vector is implied. The function $U_{pk}(q,\Omega)$ represents the irreducible vertex, and contains all information on connected averages between retarded/advanced Green's functions. 

The utility of the BS equation, is that it allows us to work in terms of only disorder averaged Green's functions. Thus, for ease of notation, we shall drop the $\langle \rangle$ brackets, and assume that all Green's functions are replaced with their disorder averaged values, unless otherwise specified. 

$\langle \hat G \rangle$ is the Green's function corresponding to the full action of Eq.~\ref{eq: S}. In order to compute this, we first separate $S$ into a part $S_0$ whose Green function is readily solvable, plus a perturbative piece $S_{\text{int}}$ that contains the disorder field. It is well known that the configurational average of a translationally invariant Green's function for such a system can then be written in the Fourier basis as the inverse of the sum of the bare Green's function pertaining to $S_0$, and a self energy operator $\hat \Sigma$, as~\cite{sheng2006introduction}

\beq
\label{eq: G def} 
\langle G_{p_\pm}^\pm(\omega_\pm)\rangle^{-1} = L^2\left(\kappa p_\pm^4 - \sigma \omega_{\pm}^2\right) - \Sigma^\pm_{p_\pm}(\omega_\pm).
\eeq
We have not yet solved for $\langle \hat G \rangle$. Instead, all effects of disorder averaging have been moved onto $\langle \hat \Sigma \rangle$. The real part of $\hat \Sigma$ renormalizes the phase velocity, while the imaginary part introduces a mean free path length, beyond which the phase information of a single monochromatic wave has been erased by scattering processes. We will find that the self-energy does, in fact, have an imaginary part, which implies that $\langle \hat G \rangle$ is a short range object, {{\em i.e.}\ its disorder average vanishes exponentially with respect to length. For the remainder of the manuscript, unless otherwise specified, we assume that we are in the weak scattering limit, whereby the imaginary part of the self energy is small.

The self energy obeys the useful relation $\hat \Sigma^+ = \left( \hat \Sigma^- \right)^*$. Using this in combination with the identity $AB = (A^{-1} - B^{-1})^{-1}(B-A)$, we can rewrite the BS equation to arrive at the Boltzmann equation~\cite{vollhardt1992self}
\beq 
\label{eq: boltzmann}
\begin{split}
\left(-\Omega  + ({ \bf v}_p \cdot { \bf q }) - \frac{1}{L^2}\frac{\Delta \Sigma_p(q)}{2 \sigma \omega } \right) \phi_{pp'}(q) = \\
 -\frac{\Delta G_p (q)}{2 \sigma \omega L^2 } \left(\delta_{pp'}+ \sum_k U_{pk}(q) \phi_{kp'}(q)\right).
 \end{split}
\eeq
The velocity 
\beq
\label{phase velocity}
{\bf v}_p = \frac{2\kappa p^2}{\sigma \omega} {\bf{p}}, 
\eeq
contains an additional term $\sim q^2 {\bf q}$ that we have discarded in anticipation of later taking the diffusive limit. 
The $\Delta$ symbol means the difference between retarded and advanced quantities. We will only be concerned with its action on the self-energy and Green's function, namely 
\begin{subequations}
\beq
\Delta \Sigma_p(q) = \Sigma^+_{p_+}(\omega_+) - \Sigma^-_{p_-}(\omega_-),
\eeq
\beq
\Delta G_p(q) = G^+_{p_+}(\omega_+) - G^-_{p_-}(\omega_-).
\eeq
\end{subequations} 
For ease of notation, when the $q$ dependence of a quantity is not explicitly written, it is evaluated at $q=0$, but not at $\Omega=0$. For example, $\Delta G_p = G_p^+(\omega+\Omega/2)-G_p^-(\omega-\Omega/2)$.

We may also consider the group velocity of undulatory waves. This depends on the self energy and can be written
\beq
\label{eq: group velocity deltas}
{\bf v}^G_p = {\bf v}_p \frac{1-\delta_1}{1+\delta_2 },
\eeq
where we have defined the dimensionless quantities
\beq
\label{eq: delta1}
\delta_1 = \frac{1}{4 \kappa p^3}\frac{\p \text{Re}\Sigma}{\p p},
\eeq
and
\beq
\label{eq: delta2}
\delta_2 = \frac{1}{2\sigma \omega}\frac{\p \text{Re}\Sigma}{\p \omega}.
\eeq
We will encounter the function $\delta_1$ frequently in our calculations. 

Inspecting Eq.~\ref{eq: boltzmann}, we see that the $p'$ index can be freely summed. Doing so allows us to work with the simpler function 
\beq
\label{eq: phi one p}
\phi_p(q) \equiv \sum_{p'} \phi_{pp'}(q).
\eeq

The right hand side of Eq.~\ref{eq: boltzmann} is proportional to $\Delta G_p$. In the weak scattering limit, the imaginary part of $\hat \Sigma$ is small and so $\Delta G_p$ as a strongly peaked Lorentzian around some value $p=p_F$ determined by the condition 
\beq
\label{eq: fermi momentum}
\text{Re}G_{p_F}^{-1}=0.
\eeq
In alignment with the terminology appearing in hard condensed matter~\cite{altland2010condensed}, we refer to this wavenumber $p_F$ as the {\it Fermi wavenumber}, though our system is purely classical. From Eq.~\ref{phase velocity}, we additionally define the associated Fermi velocity $v_F = 2 \kappa p_F^3/\sigma \omega$. The sharpness of $\Delta G_p$ around $p_F$ suggests that we may approximate it as a $\delta$-function pinning the wavenumber magnitude to $p=p_F$. Using the large $L$ limit to replace summations with integrations, we find the identity
\beq
\label{eq: delta G approximation}
\sum_p \Delta G_p(...) \xrightarrow[\text{weak scattering}]{} \frac{i \pi \nu}{\sigma \omega}\int_0^{2\pi} \frac{d\theta}{2\pi} (...)|_{p=p_F},
\eeq
which we will make extensive use of. We have further defined the density of states per unit volume
\beq
\label{eq: nu}
\nu = \frac{p_F}{2\pi v_F (1-\delta_1)}.
\eeq

Combining Eqs.~\ref{eq: BS} and~\ref{eq: delta G approximation}, we notice that $\phi_p(q)$ is strongly peaked around the forward direction, {\em i.e.}\ $\hat {\bf{ p}} \cdot \hat {\bf{q}} =0$. This suggests that $\phi_{p}(q)$ is approximately given by the first couple moments of its Legendre expansion. We define
\beq
\label{eq: S and J}
S \equiv \sum_{p} \phi_p(q), \;\;\;  {\bf J} \equiv \sum_p { \bf v}_p \phi_p(q).
\eeq

These represent the intensity propagation density $S$, and current density ${\bf J}$. The velocity ${\bf v}_p$ is the zero curvature group velocity of undulations. It is proportional to, yet not necessarily equal to, the transport velocity, {\em i.e.\,}the average velocity of intensity across the membrane. Solutions $S$ and ${\bf J}$ are found by taking the first two moments of the Boltzmann equation (Eq.~\ref{eq: boltzmann}). 

The first moment is found by summing both sides of the Boltzmann equation over $p,p'$. We obtain
\beq
\begin{split}
&-\Omega S + {\bf q} \cdot {\bf J} = \frac{\pi \nu}{i\sigma \omega L^2} + \\
&+ \frac{1}{2 \sigma \omega L^2} \sum_{pp'k}\left( \Delta \Sigma_p(q) \delta_{pk}- \Delta G_p(q) U_{pk}(q) \right)\phi_{kp'}(q).
\end{split}
\eeq
In order for $S$ to exhibit diffusive behavior, all terms $\sim S$ in the equation must vanish in the limit $q,\,\Omega \to 0$. The existence of a diffusive solution is thus contingent on the vanishing of the final term. This is indeed the case, as is ensured by the Ward identity (WI)
\beq
\label{eq: WI}
\Delta \Sigma_p(q) = \sum_{p'} U_{pp'}(q) \Delta G_{p'}(q).
\eeq
In the limit $q,\Omega \to 0$, we can replace $\Delta \Sigma_p \to 2 i \text{Im}\Sigma_p$, so that the WI relates the imaginary part of the forward scattering amplitude to the total cross section. The WI is thus a generalized optical theorem~\cite{sheng2006introduction}, and depends on the type of wave equation studied~\cite{kroha1993localization,van1991speed,van1992speed}. The derivation of the WI is nontrivial and presented in Appendix~\ref{app: WI}. 

Implementing the WI yields the continuity equation
\beq
\label{eq: continuity}
-i\Omega S + i {\bf q} \cdot {\bf J} = \frac{\pi \nu}{\sigma \omega L^2} + O(q^2).
\eeq
In position space, Eq.~\ref{eq: continuity} is of the form $\p_t S + {\bf \nabla} \cdot {\bf J} = \pi \nu/\sigma \omega L^2$, hence the name {\it continuity equation}. 

In order to obtain a closed set of hydrodynamic equations, we must relate ${\bf J}$ to S. If the intensity is to exhibit diffusive behavior, then the current ${\bf J}$ must obey Fick's law ${\bf J} = - D \nabla S$, with $D(\omega)$ some diffusion coefficient to be determined. The coefficient $D$ can, and will, depend on the internal frequency $\omega$.

We begin by taking the second moment of the Boltzmann equation, {\em i.e.}\ applying $\sum_p ({\bf v}_p \cdot {\bf \hat q}) (...)$ to both sides of Eq.~\ref{eq: boltzmann}. As we are interested in the long length/time limit, we retain only the lowest terms through $\mathcal{O}(\Omega, q)$. We obtain
\beq
\begin{split}
\label{eq: second moment}
&q\sum_p ({\bf v}_p \cdot {\bf \hat q} )^2 \phi_p(q) = \\ 
&\frac{1}{2\sigma \omega L^2} \sum_{pk} ({\bf v}_p \cdot {\bf \hat q} ) \bigg( \Delta \Sigma_p(q) \delta_{pk} -\Delta G_p(q) U_{pk}(q) \bigg)\phi_k(q) .
\end{split}
\eeq
The left hand side (LHS) is the third moment of $\phi_{p}(q)$, and prevents a closed solution in $S$ and ${\bf J}$. This is remedied in the usual way, by everywhere replacing $\phi_p(q)$ with its truncated Legendre expansion
\beq
\label{eq: phi expansion}
\phi_p(q) = \frac{\Delta G_p}{i\pi \nu /(\sigma \omega)}\left( S + \frac{2}{v_p^2}({\bf {v}}_p \cdot {\bf \hat q})({\bf \hat q} \cdot \hat {\bf J})\right).
\eeq
The LHS is evaluated using the identity $\Omega_d^{-1} \int d \Omega_d ({\bf a} \cdot {\bf \hat p})({\bf b} \cdot {\bf \hat p}) = d^{-1} {\bf a} \cdot {\bf b}$, valid for arbitrary vectors ${\bf a}$ and ${\bf b}$, in any dimension $d$, with solid angle $\Omega_d$. As a result, this term is simply 
\beq
\text{LHS}=q\frac{v_F^2}{2} S.
\eeq

The right hand side (RHS) is more difficult to simplify than it was when deriving the continuity equation, as the angular dependence in $({\bf v}_p \cdot {\bf \hat q})$ prevents direct application of the WI. Inserting the Legendre expansion of $\phi_k(q)$ and using Eq.~\ref{eq: delta G approximation} to perform the $k$ summation, we find that the term $\sim S$ on the RHS vanishes by means of the WI. Computing the remaining terms we find
\beq
\text{RHS} = \left( \frac{i \text{Im}\Sigma_{p_F}}{\sigma \omega L^2} -M_0 \right) ({\bf \hat q} \cdot {\bf J})
\eeq
where
\beq
\label{eq: M0}
M_0 = \frac{1}{i\pi \nu v_F^2 L^2} \sum_{pp'} \Delta G_p ({\bf v}_p \cdot {\bf \hat q}) U_{pp'} ({\bf v}_{p'} \cdot {\bf \hat q}) \Delta G_{p'}.
\eeq

Collecting all terms and trivially rearranging, we arrive at Fick's Law
\beq
\label{eq: ficks law}
{\bf J} = - i {\bf q} \left( \frac{L^2 \sigma \omega}{\text{Im}\Sigma_{p_F}}\frac{v_F^2/2}{1-\sigma \omega L^2 M_0/(i \text{Im}\Sigma_{p_F}(\omega))}\right) S.
\eeq
The term $M_0$ in the denominator encodes the effects of coherent scattering, and is also responsible for weak/strong localization. Eqs.~\ref{eq: continuity},~\ref{eq: phi expansion} and~\ref{eq: ficks law}, and complete the hydrodynamic description.

\section{Diffusion and weak/strong localization}
\label{sec: localization}
The hydrodynamical equations contain a wealth of information about undulatory wave propagation, whose physical meanings are opaquely hidden in $M_0$ and $\hat \Sigma$. Before directly computing the diffusion coefficient and weak localization correction, we briefly discuss computation of disorder averages in general. This has the additional benefit of laying the groundwork for analysis beyond the average intensity, for example, computing the fluctuations in intensity transport known as speckle correlations~\cite{van1999multiple}.

The well-known particle/wave duality in quantum mechanics affords a fruitful language for describing the propagation of undulatory on a membrane. A single undulatory wavepacket can be viewed as a {\it particle} that is scattered by a random {\it potential} resulting from Gauss curvature. In this language, the disorder averaged retarded (advanced) Green's function, $\langle \hat G^\pm({\bf x}, {\bf x}';t) \rangle$, gives the amplitude for one particle initially at position ${\bf x'}$ to propagate forward (backward) in time $t$ to point ${\bf x}$. In the weak scattering limit, this average can be computed by considering the path as consisting of a series of scattering events with the random potential. The perturbative series is ordered by the number of scatterings, which at a fixed distance $|{\bf x}-{\bf x'}|$ becomes smaller at weak curvature. 

In the particle formulation, $\langle \hat G^\pm \rangle$ contains only single-particle information. Scattering events are independent, and there are no interference effects between undulatory waves. In contrast, the four-point function defined in Eq.~\ref{eq: four point} contains two-particle information. It is the disorder average of two particles, one moving forward in time and one moving backward in time (called a {\it hole} or {\it anti-particle}) to the same initial/final positions. The two particles have the potential to constructively interfere with one another, which is the source of the long-range nature of the four-point function. Viewed as waves, two-particle information encodes coherence effects in the system. 

Schematically, the two particles interfere constructively when they encounter the same sequence of scattering paths. This results in a long range object for intensity transport called the {\it diffuson}. In the special case of return to the origin (${\bf x} = {\bf x'}$), time reversal invariance in the action permits another solution whereby one of the particles is traversed backwards in time. This leads to another long range object that reduces diffusion, called the {\it cooperon}. Analysis of intensity transport in terms of diffusons/cooperons is given in Appendix~\ref{app: diffuson}, where we provide a diagrammatic derivation of the hydrodynamical equations in section~\ref{sec: hydrodynamics}. The diagrammatical analysis allows one to extend beyond the level of analysis in this manuscript. In particular, one can use the formalism to describe fluctuations in intensity, which arise from diffuson-diffuson scattering~\cite{akkermans2007mesoscopic}.

All of the information regarding diffusion is contained in the long length/time limit of the four-point function, to which we now turn. By combining the continuity equation with Fick's law (Eqs.~\ref{eq: continuity},~\ref{eq: ficks law}), we arrive at
\beq
\label{eq: phi}
\phi_{pp'}(q) = \left(\frac{-1}{2 \pi \nu L^2}\right) \frac{\Delta G_p \Delta G_{p'}}{-i \Omega + D q^2},
\eeq
which has the diffusive form postulated in section~\ref{sec: signatures}. The coefficient $D$ is precisely that appearing on the righthand side of Eq.~\ref{eq: DE final}. From our hydrodynamic analysis, we further obtain
\beq
\label{eq: D inverse definition}
D^{-1} = D_0^{-1} + D_0^{-1} i \tau M_0,
\eeq

where $D_0$ is the Drude-Boltzmann diffusion coefficient~\footnote{This name was chosen to agree with the nomenclature used for this approximation in computing the conductivity of a metal. See Ref.~\cite{akkermans2007mesoscopic} for more details}}, given by
\beq
\label{eq: D0}
D_0 = \frac{1}{2} v_F^2 \tau,
\eeq
and $\tau$ is the scattering time given by 
\beq
\label{eq: tau}
\tau^{-1} = \frac{ \text{Im}\Sigma_{p_F}}{\sigma \omega L^2}.
\eeq
The diffusion coefficient of Eq.~\ref{eq: D0} takes the standard form, implying the intensity transport velocity, $v_t$, is equivalent to the group velocity evaluated at the Fermi wavenumber:
\beq
v_t = v_F.
\eeq

Both $D_0$ and $\tau$ depend only on the self-energy, which can obtained by computing the disorder average of only a single Green's function. By combining $v_t$ and $\tau$, we determine the mean free path
\beq
\label{eq: mean free path}
\ell = v_t \tau = v_F \tau.
\eeq
The scattering time and mean free path represent the average time/length before an undulation wavepacket is scattered by Gaussian curvature. In position space, this correspond to an exponential decay $\langle G^+({\bf x},0;\omega) \rangle \sim e^{-|{\bf x}|/\ell}$. 

Per Eq.~\ref{eq: D0}, $D^{-1}$ is a sum of two pieces: single-particle effects coming from $D_0$, and two-particle effects mediated via the irreducible vertex $U_{pp'}$. Setting $U_{pp'}(q)=0$ (and thereby $M_0$), is the Drude-Boltzmann approximation, whereby $\langle \hat G^+ \hat G^- \rangle$ is replaced by the product of its averages $\langle \hat G^+ \rangle \langle \hat G^- \rangle$.

In order to proceed further, we must further specify the irreducible vertex. Our arguments at the beginning of the section suggest coherent scattering will primarily lead to two effects: diffusion and weak localization. Anticipating this, we decompose $U_{pp'}(q)$ into the sum of two terms
\beq
\label{eq: U0 Umc}
U_{pp'}(q) \approx U^0_{pp'}(q) +U^{\text{(MC)}}_{pp'}(q),
\eeq
called the {\it bare} vertex $U_{pp'}^0(q)$, and the {\it maximally crossed} vertex $U_{pp'}^\text{(MC)}(q)$. These are responsible for diffusion and weak localization respectively. Likewise, we decompose the diffusion coefficient into a sum of two pieces
\beq
D^{-1} = D_{\text{c}}^{-1} + D_\times^{-1}.
\eeq
$D_c$ represents the coherent diffusion coefficient found by choosing $\hat U=\hat U^0$. Inserting $\hat U^0$ into Eq.~\ref{eq: D inverse definition} and rearranging, we identify
\beq
\label{eq: Dc}
D_c = D_0 (1+\delta_c)^{-1},
\eeq
where we have defined the reduction factor
\beq
\delta_c = i \tau M_0|_{U_{pp'}(q)=U^0_{pp'}}.
\eeq
The calculation of $\delta_c$ is presented in Appendix~\ref{app: delta c}. 

$D_\times$ is the maximally crossed diffusion coefficient found by choosing $\hat U=\hat U^\text{(MC)}$, and ignoring the $D_0^{-1}$ contribution that has already been counted. It is explicitly given by
\beq
\label{eq: D times}
D_\times^{-1} = \left. D_0^{-1} i \tau M_0 \right|_{\hat U = \hat U^\text{(MC)}}.
\eeq
We begin by first studying the $\hat U^0$ contribution to $D^{-1}$.

The bare vertex is defined as the minimally disorder averaged vertex connecting two pairs of retarded and advanced propagators. For an explicit representation of $U_{pp'}^0(q)$ in terms of diagrammatic perturbation theory (Appendix~\ref{app: self-energy}) see Fig.~\ref{fig: U0}. The bare vertex represents a single particle-hole scattering event. Inputting $U_{pp'}^0(q)$ into the BS equation generates all trajectories where the particle and hole scatter off the same sites, in the same order. In the diagrammatic derivation of Appendix~\ref{app: diffuson}, these trajectories correspond to summing over the set of all box diagrams with uncrossed disorder lines, the so-called ladder approximation~\cite{akkermans2007mesoscopic}. 

The hydrodynamic analysis has thus shown that long-range contributions to the four-point function come from summations over ladder-type diagrams, or in the position space representation, a summation over scattering events where the particle and hole traverse the same trajectory in the same order. If the system is time-reversal invariant (as is the case here) there exists an additional long-range contribution to the four-point function, found by reversing the order of scattering for one of the particles ({\em i.e.} running backwards in time). Using time-reversal symmetry, we can additionally change the signs of the hole wavenumber to obtain the identity
\beq
\phi(p_+, p'_+ ; p_-, p'_-) = \phi(p_+, p'_+; -p'_-, -p_-),
\eeq
where we have explicitly written the dependence on all four wavenumbers. Reducing to a function of only three wavenumbers we find
\beq
\phi_{pp'}(q) = \phi_{\frac{1}{2}(p-p'-q),\frac{1}{2}(p'-p-q)}(p+p').
\eeq
This identity trivially allows us to sum over all maximally crossed diagrams, by mapping them onto a summation over uncrossed diagrams. Furthermore, we know that this must be a long-range object that becomes divergent for some combination of ${\bf p},{\bf p'}$ and ${\bf q}$ as $\Omega \to 0$, and hence can have an appreciable effect.

\begin{figure}
\includegraphics[scale=0.7]{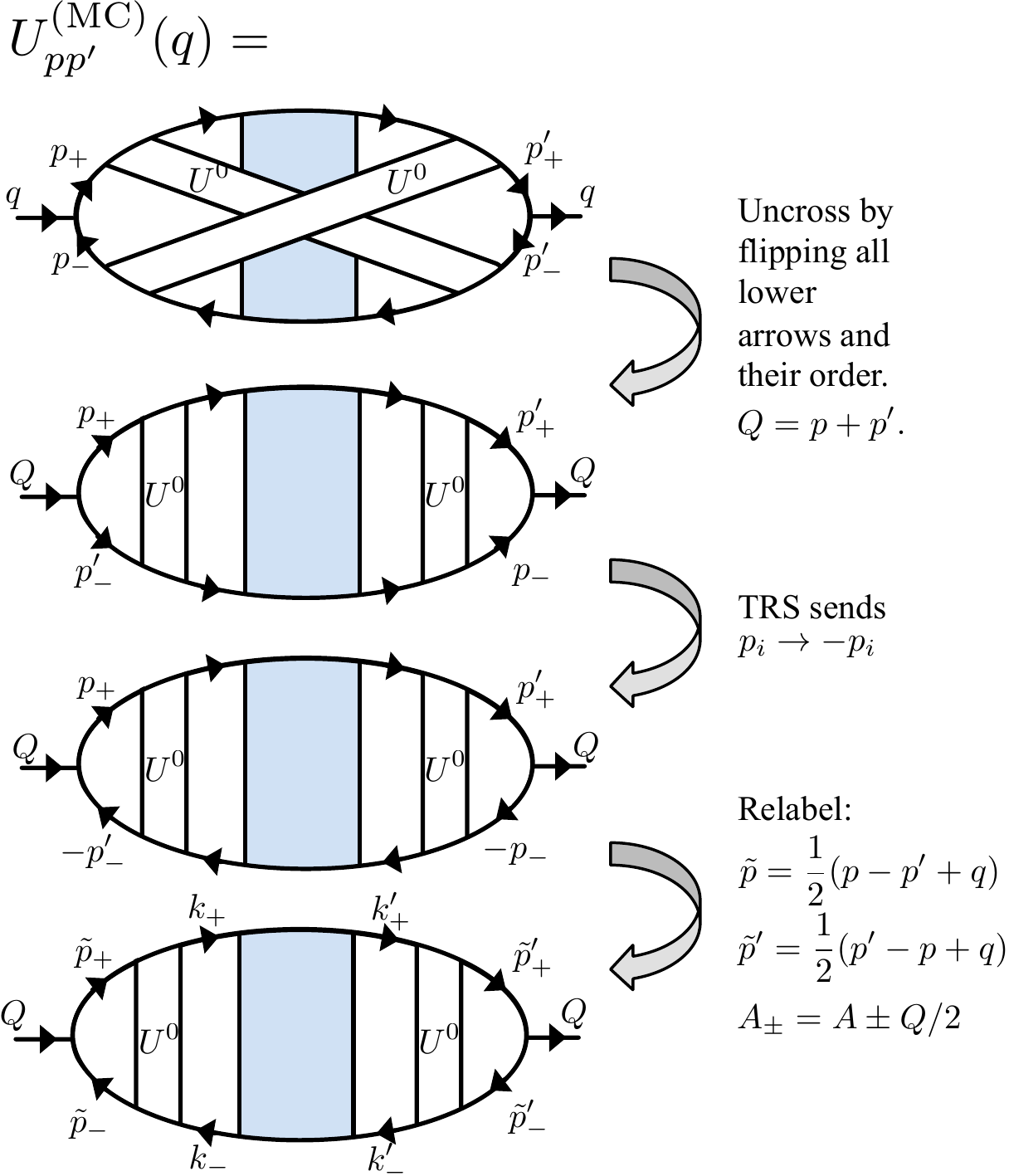}
\caption{(Color online) diagrammatic construction of the maximally crossed irreducible vertex. White boxes represent the bare irreducible vertex $\hat U^0$; the shaded box represents the diffusive four-point function in Eq.~\ref{eq: phi}. {\it Crossing} refers to overlap of $\hat U^0$ vertices. The summation over maximally crossed diagrams can alternatively be viewed as summation over uncrossed diagrams with all lines on the lower rung reversed. Time-reversal symmetry allows us to flip arrow orientation bringing it to standard form.}
\label{fig: umc}
\end{figure}

In Fig.~\ref{fig: umc}, we compute $\hat U^\text{(MC)}$ diagrammatically. The procedure is as follows: write the maximally crossed irreducible vertex by crossing two bare irreducible vertices and inserting the full four point function between them, uncross the diagram, then use time-reversal symmetry to reverse wavenumbers bringing it into a standard form. We find the equation~\cite{sheng2006introduction}

\beq
U_{pp'}^{\text{(MC)}}(q) = \sum_{k,k'} U_{\tilde pk}(Q) \phi_{kk'}(Q) U_{k' \tilde p'}(Q),
\eeq
where $\tilde p,\tilde p'$ and $Q$ are defined in Fig.~\ref{fig: umc}. As expected, $\hat U^\text{(MC)}$ diverges as $\Omega,Q \to 0$, per Eq.~\ref{eq: phi}. Working in the divergent limit where $\hat U^\text{(MC)}$ is appreciable, we discard all terms $\mathcal{O}(Q)$ and perform the summations using the WI (Eq.~\ref{eq: WI}) to obtain

\beq
U_{pp'}^{\text{(MC)}}(Q,\Omega) = \left(\frac{-1}{2\pi \nu L^2} \right)\frac{\Delta \Sigma_{\tilde p}\Delta\Sigma_{\tilde p'} }{-i \Omega + D_c Q^2}.
\eeq
The definition of $Q$ tells us that $\hat U^\text{(MC)}$ is maximal when $p' \approx -p$. In the wavenumber representation, $p$ and $p'$ correspond to the incoming and outgoing wavenumbers for an initial wave incident on the disordered media. $p \approx -p'$ thus corresponds to backscattering, which in position space corresponds to particle trajectories that return to the point of departure.

We continue the calculation of $D_\times$ by inserting $\hat U^\text{(MC)}$ into $M_0$ (Eq.~\ref{eq: M0}). We work in the $Q \to 0$ limit, approximating the summation by everywhere replacing $p'=-p$. This yields the intermediate expression
\beq
M_0 = \frac{1}{2i \pi ^2 L^4 \nu ^2 v_F^2} \sum_{p,Q} (\Delta G_p)^2 ({\bf v}_p \cdot {\bf \hat q})^2 \frac{\Delta \Sigma_{\tilde p}\Delta\Sigma_{\tilde p'} }{-i \Omega + D_c Q^2}.
\eeq
In the weak scattering approximation, we may ignore terms containing $(\hat G^\pm)^2$ inside the summation, allowing us to use $(\Delta G_p)^2 \approx -2\Delta G_p/\Delta \Sigma_p$~\cite{sheng2006introduction,akkermans2007mesoscopic}. In the diffusive limit, $q$ is also small and we can replace $\tilde p,\tilde p'$ with $p,p'$. Using Eq.~\ref{eq: delta G approximation} to perform the $p$ summation and adding a factor of $(1/2)$ from the remaining angular integration we find 
\beq
M_0 =  \frac{1}{2\pi \nu} \frac{\Delta \Sigma_{p_F}}{\sigma \omega L^4} \sum_Q \frac{1}{-i \Omega + D_c Q^2}.
\eeq
Inserting $M_0$ into Eq.~\ref{eq: D times} and using the definition of $\tau$ (Eq.~\ref{eq: tau}) we obtain the maximally crossed diffusion coefficient
\beq
D_\times^{-1}  = D_0^{-1}\frac{1}{ \pi \nu L^2} \sum_Q \frac{1}{-i \Omega + D_c Q^2},
\eeq
and from Eq.~\ref{eq: D0}, the total diffusion coefficient~\cite{vollhardt1992self}
\beq
\label{eq: diffusion coefficient equation}
D^{-1} = D_c^{-1}+D_0^{-1}\frac{1}{\pi \nu} \int \frac{d^2 Q}{(2\pi)^2} \frac{1}{-i \Omega+D_c Q^2} .
\eeq
In the last line we have taken the continuum limit. The limits of integration over $Q$ must be limited to the domain where diffusive transport is applicable. The upper bound is set by the inverse mean free path $\ell^{-1}$ defined in Eq.~\ref{eq: mean free path}, while the lower bound is set by the inverse system size $L^{-1}$. 

In the weak disorder limit, $D \approx D_c$, and so the rightmost term of Eq.~\ref{eq: diffusion coefficient equation} may be considered small. Taking the long time $\Omega \to 0$ limit, performing the integration, and using the definition of $\nu$ in Eq.~\ref{eq: nu} we find
\beq
D= D_c - \delta D,
\eeq
where we have defined the weak localization correction
\beq
\label{eq: weak localization}
\delta D = \frac{2 \kappa p_F^2(1-\delta_1)}{\pi \sigma \omega (1+\delta_c)} \ln\left(\frac{L}{\ell}\right).
\eeq

This is one of our main results. The summation of maximally crossed diagrams has led to a reduction in the diffusion coefficien, which diverges like $\ln L$ as $L \to \infty$. The logarithmic dependence on system size is a universal result for 2D mesoscopic systems~\cite{akkermans2007mesoscopic}, and appears in our model as well. Formally, in the $L \to \infty$ limit all states are localized, however, since the weak localization correction is only logarithmic, if may be difficult for finite systems to localize.

All of the quantities appearing in $\delta D$ have been computed elsewhere already ($\ell$, $p_F$, and $\delta_1$ in Appendix~\ref{app: self-energy}, $\delta_c$ in Appendix~\ref{app: delta c}), and can be determined immediately. In Appendix~\ref{app: full expressions} we give the full analytic expressions in $d_H=0,1,2$ for $\delta D$, as well as the other main quantities in this manuscript.

The weak localization correction breaks down near the localization transition ($\delta D \approx D_c$). In order to probe the onset of localization, we treat Eq.~\ref{eq: diffusion coefficient equation} self-consistently, replacing $D_c$ under the integration on the right side with the full diffusion coefficient $D$~\cite{vollhardt1992self}. This has the effect of renormalizing each of the two-particle propagators appearing in the ladder diagrams of $D_c$ with maximally crossed vertices, and vice versa for those appearing in $D_\times$. In the diagrammatic language, this corresponds to renormalizing the diffuson with all numbers of insertions of cooperons. In the localized regime, we expect the diffusion coefficient to vanish at long times. We can then posit a localization length~\cite{vollhardt1992self,sheng2006introduction}
\beq
\label{eq: xi def}
\xi^2 = \lim_{\Omega\to0}\frac{D(\Omega)}{-i \Omega}.
\eeq
The localization length $\xi$ determines the length beyond which all states are localized. Assuming such a length exists in the system, we no longer need to restrict the wavenumber integration over Q to be larger than $L^{-1}$, as $\xi$ will regulate the low wavenumber divergence. Solving the self-consistent extension of Eq.~\ref{eq: diffusion coefficient equation} we find
\beq
\label{eq: xi}
\xi/\ell= \left( e^{4 \pi^2 \nu D_0} -1 \right)^{1/2}.
\eeq
The localization length describes transport in the strong scattering regime. This result, combined with the weak localization solution of Eq.~\ref{eq: weak localization}, and the coherent diffusion coefficient $D_c$ of Eq.~\ref{eq: Dc}, completes our description of undulatory transport on a warped membrane.

\section{Results}
\label{sec: results}
\begin{table*}
\caption{Dimensionless scattering time $\tau = \omega/\text{Im}\overline{\Sigma}_{p_F}$, and dimensionless transport velocity $v_t$, in each of $d_H=0,1,2$. These are calculated within the SCSA (see Appendix~\ref{app: self-energy} for more details). We have further used the weak-scattering approximation to determine $\text{Im}\overline{\Sigma}_{p_F}$, and hence $\tau$. For $d_H<2$, the continuum picture breaks down and quantities depend on the underlying lattice. This is taken into account by restricting wavenumbers to lie below an upper cutoff $\Lambda \sim 1/a$, for $a$ the lattice spacing.}
\begin{ruledtabular}
\begin{tabular}{l c c}
 $$
 &$\tau$
 &$v_t$
\\ \hline
 $d_H=0:$
 &$\left(\frac{3}{2}\right)^{2/3} \frac{\Lambda^{4/3}}{\pi  \omega ^2 \alpha^{1/3}}$
 &$2\sqrt{\omega} \sqrt{\frac{2}{3\alpha  \Lambda ^2}}$
  \\
 $d_H=1:$
 &$\frac{6}{19 \pi \alpha^2 \omega} (1 + \frac{3\alpha/2 + 12 \alpha \ln(\Lambda/\sqrt{\omega})+1}{\sqrt{3\alpha + 12 \alpha \ln(\Lambda/\sqrt{\omega})+1}}  )$
 &$2\sqrt{\omega}\left(\frac{2}{1+\sqrt{12 \alpha  \ln \left(\frac{\Lambda}{\sqrt{\omega }}\right)+3 \alpha +1}}\right)^{3/4}$
 \\
$d_H=2:$
 &$ \frac{2 \omega}{3\pi \alpha^2}\left(1 +\frac{3}{2} \ln(1+\frac{\alpha}{\omega})\right)$
 &$2 \sqrt{\omega } \left(\sqrt{\frac{9 \alpha ^2}{4 \omega ^2}+1}-\frac{3 \alpha }{2 \omega }\right)^{3/2}$\footnote{$\alpha \ll \omega$}, \, $2\sqrt{\omega}\left(\frac{2 \sqrt{2} }{\left(6 \ln \left(\frac{\alpha }{\omega }\right)+7\right)^{3/4}}\right)$\footnote{$\alpha \gg \omega$}
\end{tabular}
\end{ruledtabular}
\label{tab: one particle data} 
\end{table*}

There are two fundamental quantities to compute: the self-energy $\hat \Sigma$ and the coherent diffusion reduction factor $\delta_c$, from which, all other quantities may be derived. Calculation of these quantities must be done starting directly from the DMV equations for physical membranes, Eqs.~\ref{eq: DMV1} and~\ref{eq: DMV2}, or the action in Eq.~\ref{eq: S}, if working with generalized membranes. Computation of $\hat \Sigma$ is lengthy, so to simplify matters we shall immediately switch to working with dimensionless quantities.

We measure lengths in units of $\sqrt{\kappa/Y}$ and time in units of $\sqrt{\kappa \sigma}/Y$. Henceforth, we redefine wavenumber and frequency
\begin{subequations}
\label{eq: dimless}
\beq
\label{eq: p bar}
p \to p/\sqrt{\kappa/Y},
\eeq
\beq
\label{eq: omega bar}
\omega \to \omega/\sqrt {\kappa \sigma} /Y,
\eeq
\end{subequations}
so that $p,\,\omega$ refer to the dimensionless wavenumber and frequency. These, in turn, lead us to define the dimensionless self energy
\beq
\label{eq: dimsig}
\bar \Sigma_p = \frac{\Sigma_p}{\kappa L^2 (Y/\kappa)^2},
\eeq
and disorder amplitude
\beq
\label{eq: alpha}
\alpha = \frac{\gamma}{16 \pi} \left( \frac{Y}{\kappa}\right)^{2-d_H}.
\eeq
This choice of length and time conveniently allows us to set $Y,\kappa,\sigma=1$ in the DMV equations.

From Eq.~\ref{eq: G def}, the self-energy is found via a disorder average of the DMV Green's function. For any given realization of the quenched background height field, the DMV equations represent a well-defined, albeit intractable, pair of partial differential equations. To overcome this, one typically decomposes the Green's function into a zero disorder contribution with known solution, plus corrections that tend to zero as the amplitude of the quenched height field vanishes. 

In Appendix~\ref{app: perturbation series} we compute the first order correction to $\overline{\Sigma}_p$. We find that the perturbation series is governed by a $d_H$-dependent parameter $Y h_\text{eff}^2(p)/\kappa$ (defined in Eq.~\ref{eq: heff})~\cite{kovsmrlj2013mechanical}, that is system size dependent for $d_H \leq 1$ and divergent at $p=0$ for $d_H > 1$. The perturbation series fails, and any perturbative computation must include a partial re-summation of some set of terms.
  
To remedy the failure of the perturbation series, we employ the self-consistent screening approximation (SCSA)~\cite{radzihovsky1992scsa,radzihovsky1992crumpled,Zakharchenko2010,Gazit2009}. In terms of generalized membrane with two internal dimensions and $d_c$ normals, the SCSA provides the leading order term in a $d_c^{-1}$ expansion of the self-energy. The SCSA has found success in determining the renormalized elastic constants of both athermal~\cite{kovsmrlj2013mechanical} and thermal~\cite{kovsmrlj2014thermal} warped membranes.

The self-energy is computed in Appendix~\ref{app: self-energy}, where we also develop the diagrammatic perturbation theory. The remaining quantity to calculate, the reduction coefficient $\delta_c$, is computed in Appendix~\ref{app: delta c}. We hereafter discuss only the results, and refer the reader to the appropriate appendix for further details.

Beginning with one-particle quantities, in Tab.~\ref{tab: one particle data} we present results for $v_t$ and $\tau$ in $d_H=0,1,2$. The Drude-Boltzmann diffusion coefficient and mean free path can easily be found from these two quantities using Eqs.~\ref{eq: D0} and~\ref{eq: mean free path} respectively.

For membranes that are flat in the stress-free state, the transport velocity is equal to the group velocity $2\sqrt{\omega}$. The frequency dependence is a consequence of the biharmonic $\nabla^4$ term appearing in the DMV equations. For all values $d_H$, and at fixed frequency, both the transport velocity and scattering time decrease due to geometrical disorder. The former is a consequence of random scattering, which prevents undulatory waves from propagating along a straight line. The latter is due to an increased density of scatters with increasing disorder. Generally, the slowing down of waves due to multiple scattering gives rise to diffusive transport.

In Fig.~\ref{tab: asymptotics}, we give asymptotic results at small/large $\alpha$ for the mean free path $\ell$. We find at small $\alpha$, that $\ell \sim \alpha^{-5/6}$ in $d_H=0$ and $\alpha \sim \alpha^{-2}$ for both $d_H=1,2$. As expected, increased disorder leads to a reduction in $\ell$, and as a result, phase information of an undulatory wave is lost at shorter distances from the point of force application. 

The frequency dependence of $\ell$ is more interesting. For $d_H =2$, $\ell \sim \omega^{3/2}$ increases with frequency, while for $d_H \neq 2$ it decreases. Waves with high frequency can better resolve the geometry of the surface, as their characteristic wavelengths are smaller. One would expect that at higher frequency the effective curvature is smaller, leading to fewer scattering events and a longer mean free path. 

The breakdown of this explanation for $d_H<2$ is due to the lack of a well-defined curvature, which depends on two spatial derivatives of the quenched height field. There is no meaningful derivative that can be assigned to the quenched height field, as the derivatives of the height field become arbitrarily large as the lattice spacing tends to zero. The continuum picture breaks down and quantities depend on the underlying lattice spacing $a$. This is taken into account by restricting wavenumbers to lie below an upper cutoff $\Lambda \sim 1/a$ (this is an inverse length, and per Eq.~\ref{eq: p bar}, written in units of $\sqrt{\kappa/Y}$). The decrease in mean free path with respect to frequency for $d_H\leq1$ can be understood as the wave scattering off the now-resolved short-distance roughness, which would otherwise be smoothed over.

$d_H=1$ is the marginal case; the system develops logarithm dependence on $\Lambda$. The $d_H=0$ case (white noise disorder) however, is more extreme. We find that $\Lambda$ dominates the behavior of the system; indeed, from Tab.~\ref{tab: asymptotics}, the coherent diffusion coefficient $D_c \sim (\alpha^2 \Lambda)^{-2/3}$. Though we have studied the properties of the $d_H=0$ membrane (and list the corresponding results), we shall restrict our analysis to the more physical cases of $d_H=1,2$. We now consider two-particle quantities, {\em i.e.} the diffusion coefficient and localization length.

\begin{table*}
\caption{Asymptotic limits of the primary quantities contributing to localization: the coherent diffusion coefficient $D_c$, the weak localization correction $\delta D$, and the mean free path $\ell$, in each of $d_H=0,1,2$.}
\begin{ruledtabular}
\begin{tabular}{l ccc ccc}
 &\multicolumn{3}{c}{$\alpha \ll 1$}&\multicolumn{3}{c}{$\alpha \gg 1$}\\
 & $D_c$  &$\delta D$\footnote{The weak localization correction $\delta D$ is only defined for $L > \ell$, {\em i.e.} when the argument of $\ln$ is greater than one.} &$\ell$
 & $D_c$  & $\delta D$\footnotemark[1] & $\ell$
\\ \hline
 $d_H=0:$ 
 	&$\frac{2}{\pi}\frac{(2/3)^{1/3}}{\alpha^{4/3}\omega \Lambda^{2/3}}$
	&$\frac{1}{\pi}\left(\frac{3}{2}\right)^{4/3}\alpha^{1/3}\Lambda^{2/3}\ln\left(\frac{L}{\ell}\right)$
	&$ \frac{(96)^{1/6}}{\pi} \frac{\Lambda^{1/3}}{\alpha^{5/6}\omega^{3/2}}$
	&-\footnote{These are the same as the $\alpha \ll 1$ limit, as $\Lambda$ is the dominant parameter.}
	&-\footnotemark[2]
	&-\footnotemark[2]
	\\
 $d_H=1:$  
 	& $\frac{48}{(11+76 \pi ) \alpha ^2}$
	&$\left(\frac{152}{11+76 \pi }+\mathcal{O}(\alpha \ln \Lambda)\right)\ln\left(\frac{L}{\ell}\right)$
	&$\frac{12}{19 \pi  \alpha ^2 \sqrt{\omega }}$
	&$\sim \alpha^{-9/4} \ln^{-1/4}\left(\frac{\Lambda}{\sqrt{\omega}}\right)$
	&$\sim \frac{\ln (L/\ell)}{\alpha^{1/8}\ln^{1/8}(\Lambda/\sqrt{\omega})}$
	&$\sim \frac{\ln^{5/8}(\Lambda/\sqrt{\omega})}{\alpha^{11/8} \sqrt{\omega}}$
	\\
 $d_H=2:$\footnote{The precise limits here are instead $(\alpha/\omega)\ll 1$ and $(\alpha/\omega) \gg 1$} 
 	&$\frac{16 \omega^2}{3 (1+4 \pi ) \alpha^2}$
	&$\frac{8}{1+4 \pi } \ln\left(\frac{L}{\ell}\right)$
	&$\frac{4\omega^{3/2}}{3\pi\alpha^2}$
	&$ \sim \frac{\omega^2}{\alpha^2}$
	&$\sim \ln\left(\frac{L}{\ell}\right)$
	&$\sim \frac{\omega^{3/2}}{\alpha^2}$
\end{tabular}
\end{ruledtabular}
\label{tab: asymptotics} 
\end{table*}

In Appendix.~\ref{app: full expressions}, we list the full analytic expressions for $D_c, \, \delta D, \, \xi, \, \ell$ and $\nu$, which in combination with the contents of Tab.~\ref{tab: one particle data}, comprise the main results of our manuscript. In Tab.~\ref{tab: asymptotics} we give asymptotic limits at small/large $\alpha$ of the quantities of interest, namely the coherent diffusion coefficient and weak localization correction.

\begin{figure}
\includegraphics[scale=0.6]{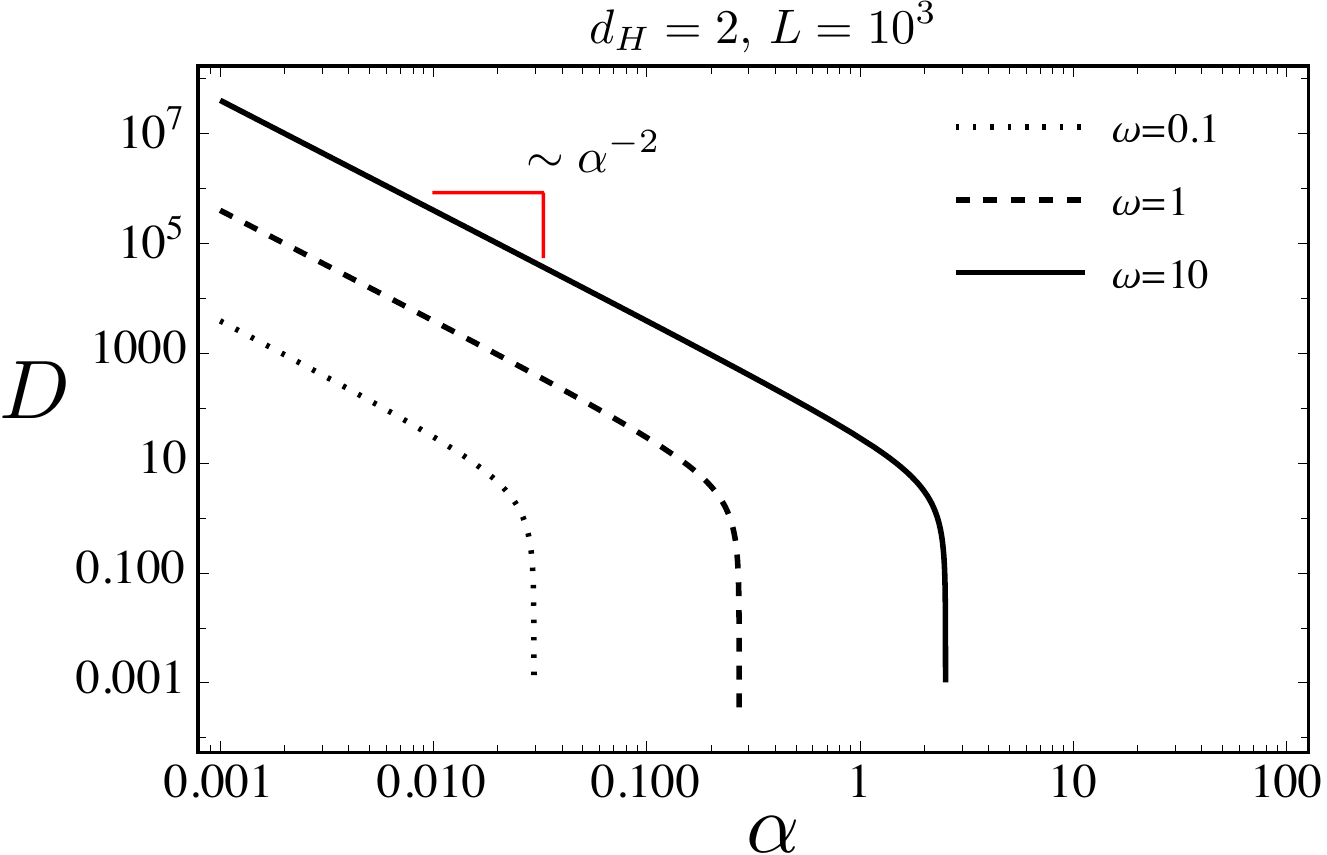}
\includegraphics[scale=0.6]{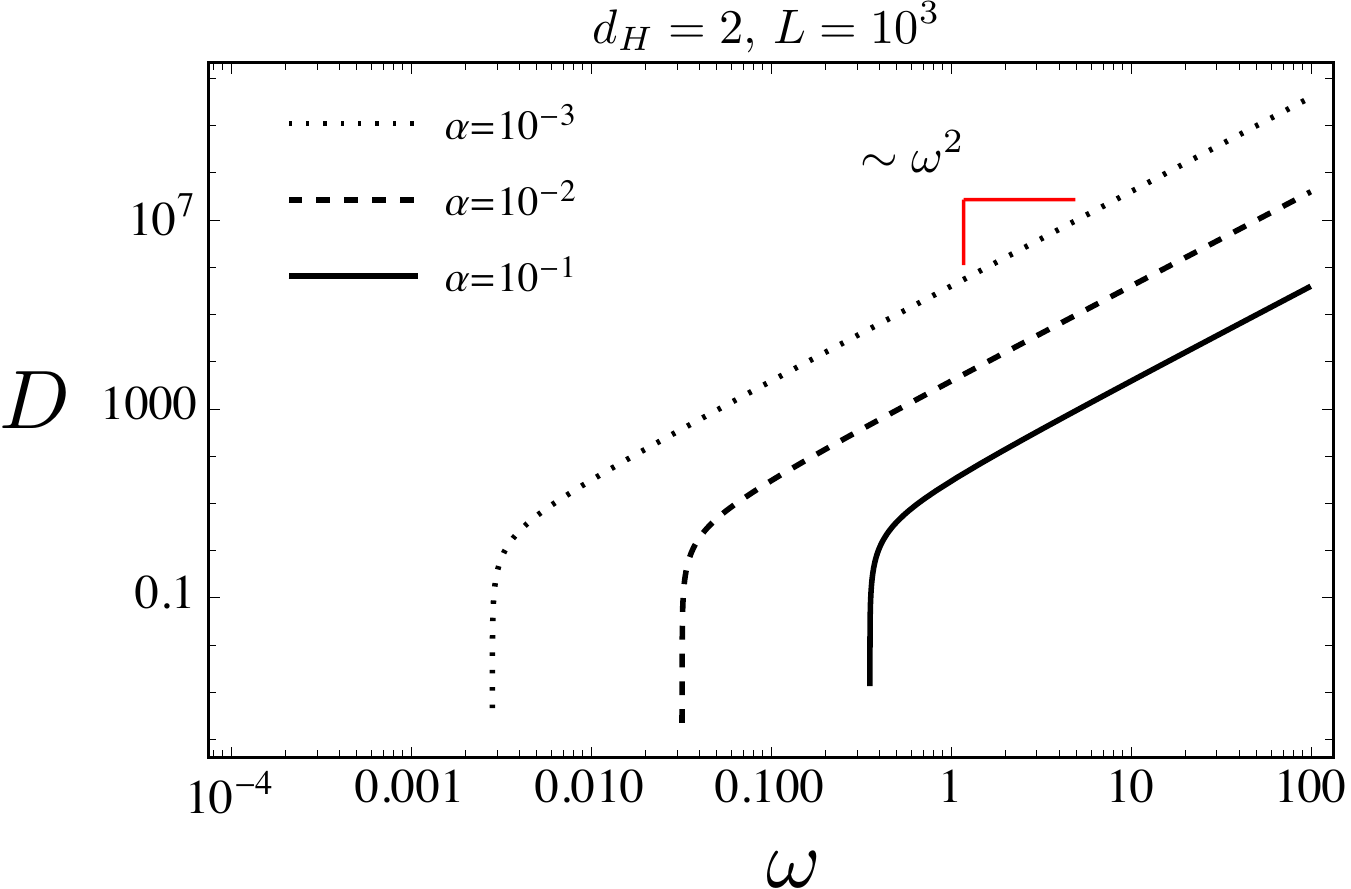}
\caption{(Color online) diffusion coefficient as a function of both disorder $\alpha$ (top) and frequency $\omega$ (bottom). $D$ decays $\sim \alpha^2$ at small $\alpha$, and grows like $\omega^2$ at large $\alpha$. At low $\omega$ there is a localization transition where $D$ vanishes. At smaller system size $L$, the localization transition is pushed to smaller/larger frequencies/disorder.}
\label{fig: D dh2}
\end{figure}

Beginning with $d_H=2$, in Fig.~\ref{fig: D dh2} we plot the total diffusion coefficient $D$ as a function of both $\alpha$ and $\omega$. $D$ depends only on the ratio $\alpha/\omega$, and so its frequency dependence is trivially inverse to that of its disorder dependence. Focusing on the case of fixed disorder, we see that at low $\omega$ the diffusion coefficient sharply drops to zero. This occurs near $D_c \approx \delta D$ and corresponds to the localization transition. The localization transition frequency grows linearly with $\alpha$, telling us that as disorder increases low frequency states are first to localize. In the weak disorder regime, we find that $D$ decreases $\sim\alpha^{-2}$ and, conversely, at fixed $\alpha$ increases $\sim \omega^2$. The increase of $D$ with $\omega$ indicates disorder acts like a high-pass filter, trapping low frequency undulatory waves.

\begin{figure}
\includegraphics[scale=0.6]{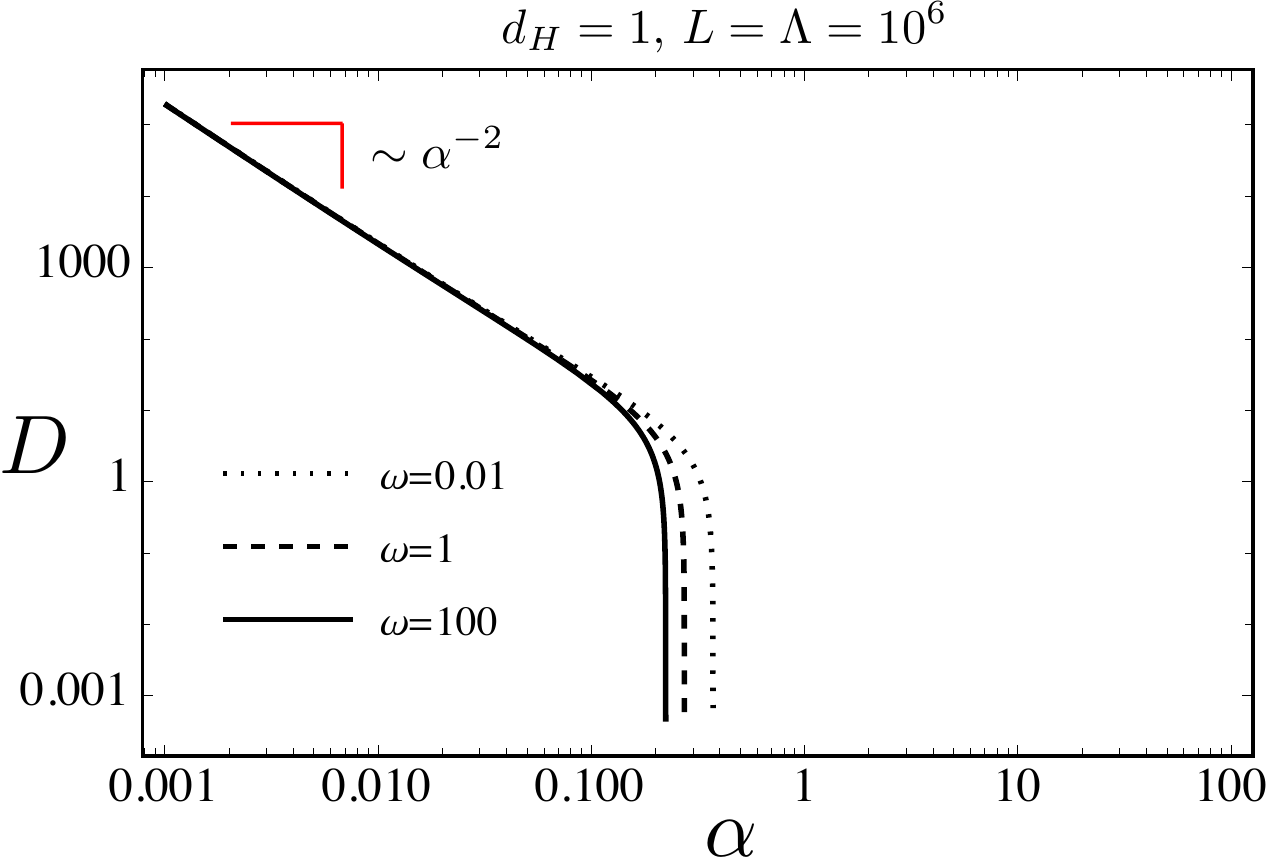}
\includegraphics[scale=0.6]{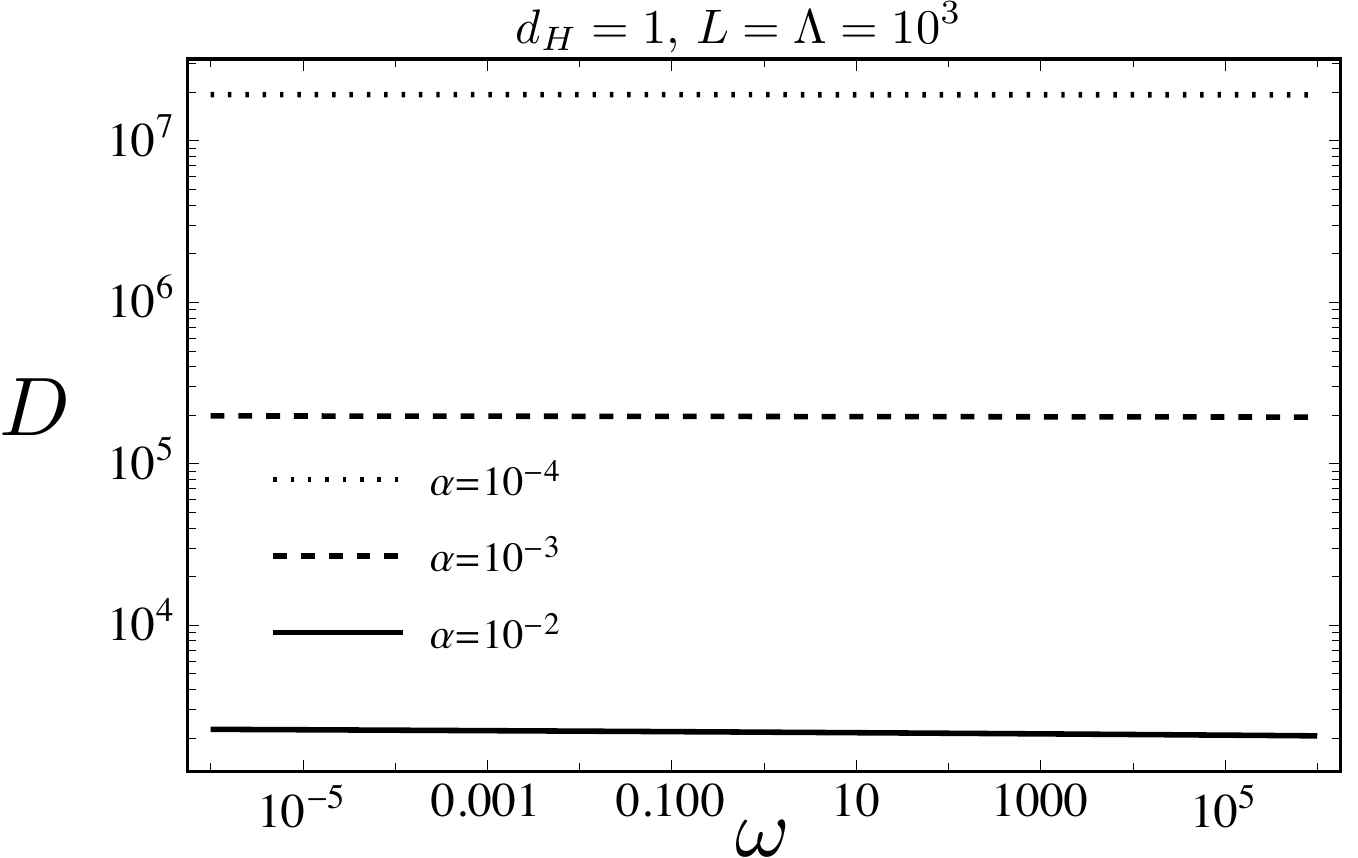}
\caption{(Color online) diffusion coefficient in $d_H=1$ versus both disorder amplitude (top) and frequency (bottom). As for $d_H=2$, $D\sim \alpha^{-2}$ at low alpha, and reaches a localization transition at large $\alpha$. The transition decrease logarithmically in frequency. If the system localizes, it occurs first at high frequency. In the lower panel, the weak frequency dependence of $D$ is shown to hold ten orders of magnitude, and even up to the upper limit $\omega \approx \Lambda^2$.}
\label{fig: D dh1}
\end{figure}

\begin{figure}
\includegraphics[scale=0.6]{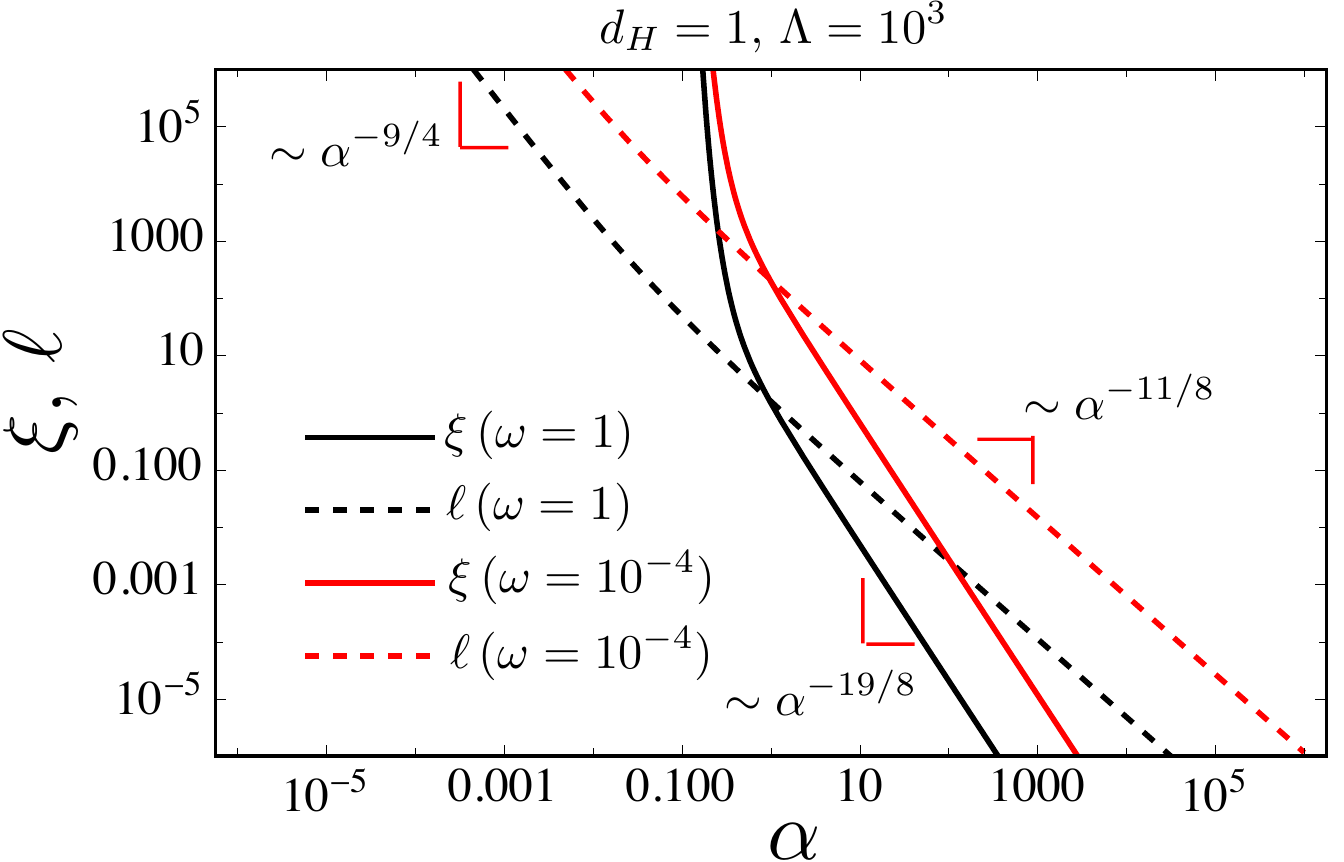}
\includegraphics[scale=0.6]{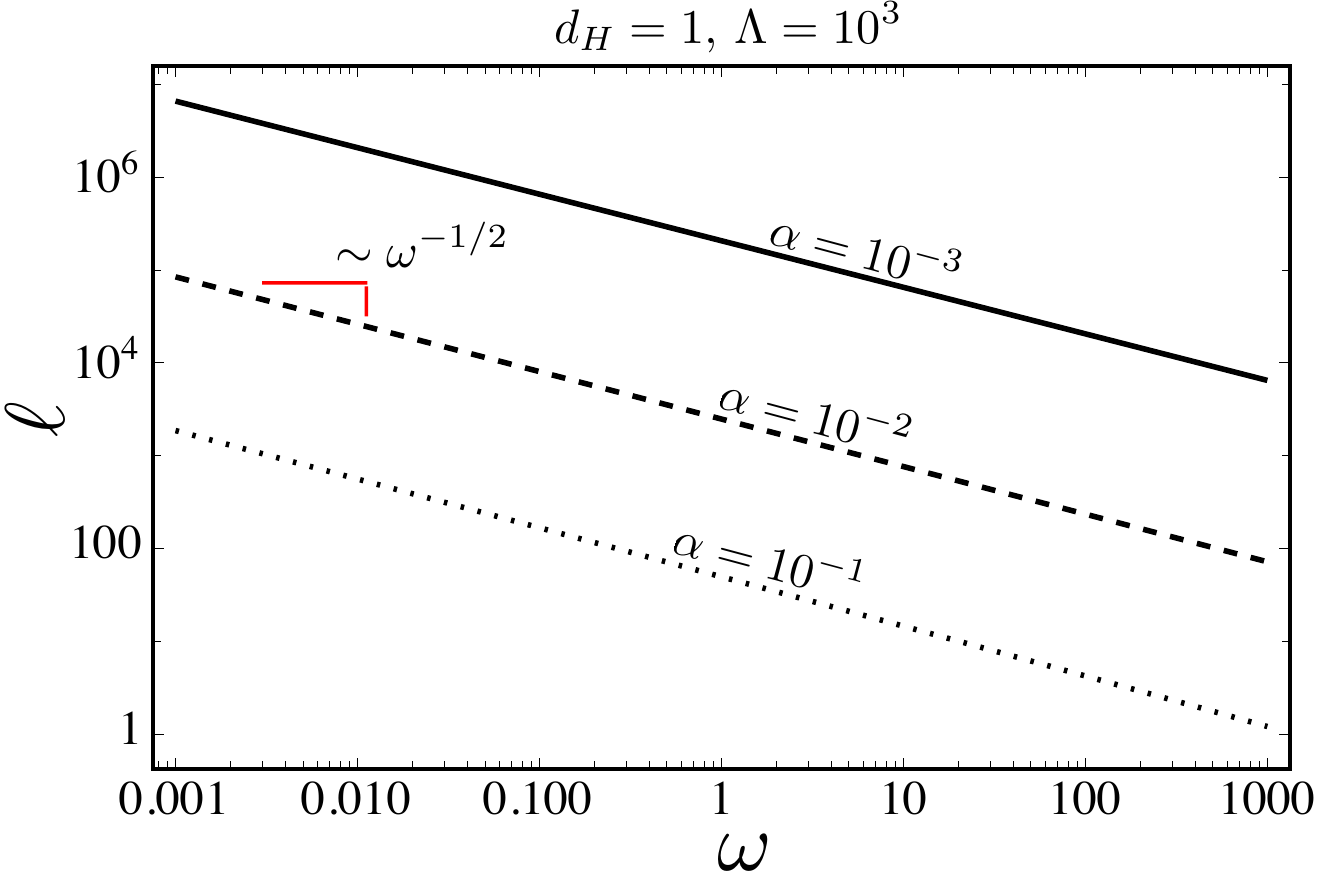}
\caption{(Color online) localization length $\xi$ and mean free path $\ell$ in $d_H=1$ as a function of both disorder amplitude (top) and frequency (bottom). In the top panel, black (red) lines correspond to $\omega=1$ ($\omega=10^{-4}$). Solid (dashed) lines refer to $\xi$ ($\ell$). The wave localizes when $\xi$ falls below $\ell$. This occurs at weaker disorder (smaller $\alpha$) with increasing frequency. In the lower panel, we show the $\omega^{-1/2}$ decay of the mean free path. The localization length is $\omega$-independent, and not shown.}
\label{fig: xi dh1}
\end{figure}

In Fig.~\ref{fig: D dh1}, we plot the $d_H=1$ diffusion coefficient versus both $\alpha$ and $\omega$. As a function of $\alpha$, $d_H=1$ behaves similarly to $d_H=2$. $D \sim \alpha^{-2}$ at low $\alpha$, and there is a localization transition at high disorder. In contrast, $d_H=1$ has only logarithmic frequency dependence. Increasing the frequency by orders of magnitude marginally decreases the value $\alpha$ at which the wave localizes. The logarithmic frequency dependence of $D$ is shown clearly in the lower panel of Fig.~\ref{fig: D dh1}, where in a log-log plot it appears as a flat line. At fixed $\alpha$, increasing the frequency over several orders of ten orders of magnitude does not significantly alter $D$, even as $\omega$ approaches its upper limit $\Lambda^2$. $D$ is not entirely independent of frequency however, as in the upper panel of Fig.~\ref{fig: D dh1} we see that increasing $\omega$ lowers the disorder amplitude at the localization transition. Though the transition point only decreases logarithmically with frequency, this behavior is still in contrast to that observed in $d_H=2$, where increasing frequency raises the localization transition disorder amplitude. This analysis at large $\alpha$ is circumspect however, as the calculation of $D$ assumes we are in the weak scattering regime. Despite this, a strong scattering calculation of the localization length (shown in Fig.~\ref{fig: xi dh1}) confirms that for $d_H=1$, high frequency waves are first to localize. 

\begin{figure}
\includegraphics[scale=0.6]{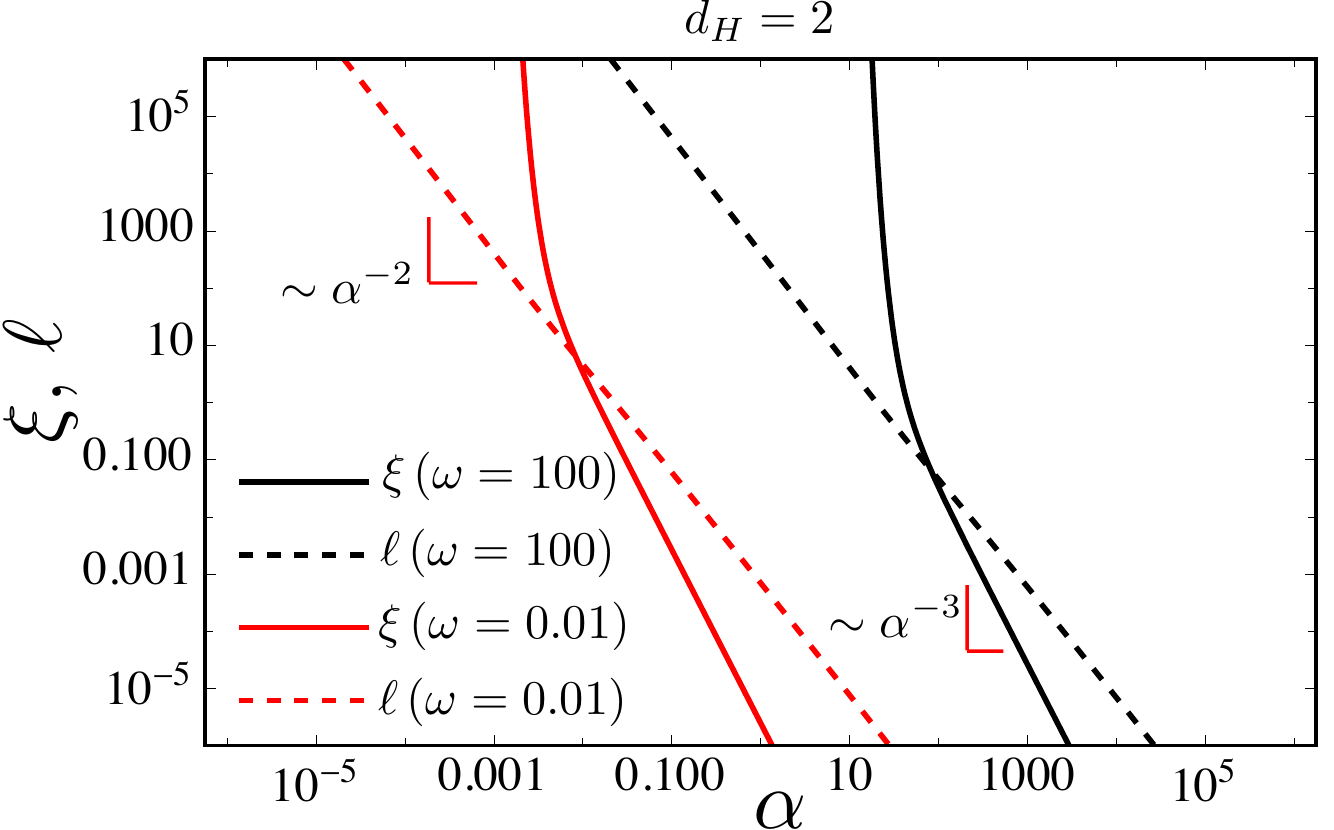}
\includegraphics[scale=0.6]{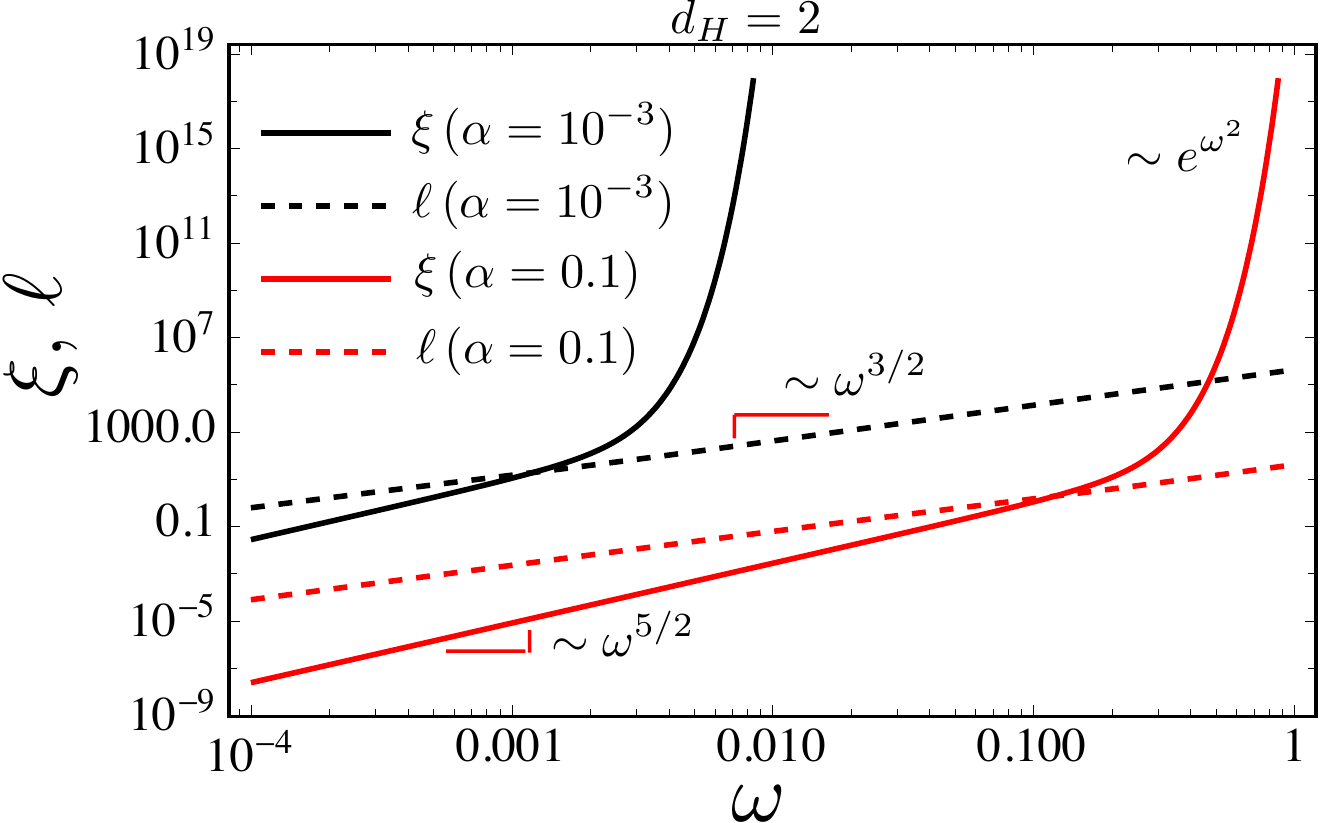}
\caption{(Color online) top: localization length and mean free path as a function of disorder in $d_H=2$. Solid (dashed) lines refer to $\xi$ ($\ell$). Black is for $\omega=10^{-3}$, red is for $\omega=0.1$. Bottom: $\xi$ and $\ell$ as a function of frequency at fixed $\alpha=10^{-3}$ (black) and $\alpha=0.1$ (red). Localization occurs at both high disorder and low frequency.}
\label{fig: xi dh2}
\end{figure}

The computation of $D$ is only valid in the weak scattering approximation, $\alpha \ll 1$. The vanishing of $D$ in both $d_H=1,2$ above a certain value of $\alpha$ signals a transition to the localization regime. To probe this, we consider the localization length $\xi$ defined in Eq.~\ref{eq: xi}, and its size relative to the mean free path $\ell$.

In Fig.~\ref{fig: xi dh2} we plot $\xi$ and $\ell$ as functions of both $\alpha$ and $\omega$ for $d_H=2$. Localization occurs approximately when $\xi<\ell$: in other words, when the wave has not yet had a chance to scatter before being localized. In agreement with the high $\alpha$ prediction of Fig.~\ref{fig: D dh2}, at large enough disorder, the localization length becomes shorter than the mean free path. When frequency is decreased, the the disorder amplitude at the localization transition decreases as well. The lower panel of Fig.~\ref{fig: xi dh2}, shows that $\xi$ transitions from $\sim \omega^{5/2}$ to $~\sim e^{\omega^2}$ dependence near the transition. The exponential increase of localization length tells us that undulatory waves are sharply divided between extended and localized.

In Fig.~\ref{fig: xi dh1}, we perform the same analysis of $\xi$ and $\ell$ in $d_H=1$ as we did for $d_H=2$. As expected, the localization length decays with increasing disorder ($\sim \alpha^{-19/8}$) faster than the mean free path at both small ($\sim \alpha^{-9/4}$ and large $(\sim \alpha^{-11/8})$ disorder. In contrast to $d_H=2$, the ratio $\xi/\ell$ is frequency independent in $d_H=1$. The lower panel of Fig.~\ref{fig: xi dh1} shows the mean free path is a decreasing function of frequency ($\sim \omega^{-1/2}$). At fixed disorder, $\xi \sim \ell$ and so the localization length will run parallel to the mean free path. Frequency does not affect the localization transition.

Though we can take the strong scattering limit via our self-consistent treatment of $D$ leading to $\xi$, these results should be interpreted with caution for two reasons. First, at high $\alpha$, the quenched height field may produce stronger curvature. Our derivation of the warped membrane equations assume that derivatives $|\nabla h|^2$ were small, thus allowing us to use the flat metric. A proper extension to strong curvature would require the use of covariant derivatives and a non-flat metric, significantly increasing the difficulty. Second, the underlying DMV equations assume that the characteristic wavelength of deformations is much smaller than the radius of curvature (see discussion following Eq.~\ref{Displacement vector}. When this does not hold, there are additional contributions to the change in the curvature tensor ${\bf K}_{ab}$ (see the definition in Eq.~\ref{linear strain}) that couple stretching to bending~\cite{niordson2012shell}.

\section{conclusion}

We analyze the transport of undulatory waves on a membrane whose stress-free state is curved. Due to Gauss' {\it theorema egregium}, curvature couples in-plane stretching deformations to the much softer our-of-plane bending undulations. As a result, undulatory wave are scattered by changes in Gauss curvature, a purely geometric mechanism. 

We study a particular class of curved surfaces called warped membranes~\cite{kovsmrlj2013mechanical}. The stress-free state of these membranes is nearly flat, and can be parametrized by a quenched height field $h({\bf x})$ drawn from a Gaussian ensemble with variance set by Eq.~\ref{disorder correlation}. By considering ensemble averages over membrane realizations, we can compute general transport quantities, independent of a particular realization of disorder.

The elastic equations of a nearly flat membrane are described using the linearized DMV equations. This isolates the effects of curvature, which can be appreciable before nonlinearities need be accounted for. Typically, membranes are much stiffer to stretching than bending, and so the in-plane deformations relax on a time scale much shorter than undulations. Integrating out these in-plane modes, we arrive at an effective, linear field theory of undulatory waves. The resulting undulatory equations of motion appear similar to those describing wave propagation in random media, albeit with a complicated nonlocal potential and biharmonic term.

As undulations obey a linear partial differential equation, the amplitude of undulations in response to a transverse applied load can be described by a Green's function. We consider an experiment whereby an initially undeformed membrane is suddenly plucked at the origin, thereby injecting energy into the system. On average, the amplitude at any point is the sum of many random phases, and averages to zero. The disorder averaged Green's function is thus a short range quantity. In contrast, the energy is a conserved quantity, and it must survive disorder averaging. Its propagation through (or lack thereof) the membrane is an indicator for diffusion/localization. We find that the kinetic energy alone, and more specifically the undulation intensity (a product of a retarded and advanced Green's function), is a long-range object capable of describing diffusion/localization. We focus on studying undulation intensity transport in response to an applied transverse load of fixed frequency at the origin. This defines a frequency dependent diffusion coefficient $D(\omega)$, from which we can determine the diffusion coefficient of any finite sized wavepacket.

Our theory obeys a Ward identity (WI) relating the self-energy to the irreducible four-point function. Using the WI, we derive diffusive dynamics of undulation intensity for long times and lengths. The resulting expression for the diffusion coefficient is found to depend critically on the type of noise under consideration, as well as the frequency of the applied load. For Gaussian noise with power law variance in wavenumber space, we find that only membranes with variance $ \langle |h_q|^2 \rangle \sim q^{-2 d_H}$ with $d_H > 1$ are well defined and independent of the short distance cutoff, with $d_H=1$ the marginal case.

For all types of noise, geometry is found to decrease both the transport velocity and scattering time. The strength of the effect increases with amplitude of the quenched height field. The combination of random scattering and slowed propagation is responsible for diffusive behavior of disorder averaged intensity transport.

Considering the effect of coherent scattering on intensity transport, we compute the diffusion coefficient and its weak localization correction in the limit of weak scattering (small quenched height field amplitude). The weak localization correction is found to behave similar to those in other 2D systems, lowering the diffusion coefficient $\sim \ln L$, which is logarithmically dependent on the system size.~\cite{sheng2006introduction,akkermans2007mesoscopic,vollhardt1992self} 

For membranes belonging to the $d_H=2$ ensemble, we find at fixed frequency the diffusion coefficient decrease like $\alpha^{-2}$, for $\alpha$ the dimensionless amplitude of the quenched height field. At large enough $\alpha$, the system undergoes a localization transition, whereby the diffusion coefficient vanishes. The weak scattering prediction is confirmed by self-consistently extending the weak localization correction to the strong scattering regime, where we find the localization length $\xi$ to decrease with $\alpha$. At fixed $\alpha$, waves first localize at low frequency. Away from the localization transition, the diffusion coefficient grows $\sim \omega^2$ with increasing frequency. This effects of random geometry are mitigated at high frequency.

For $d_H=1$, the diffusion coefficient decreases $\sim \alpha^{-2}$ until a localization transition at high $\alpha$, just like for $d_H=2$ membranes. However, for $d_H=1$, both the diffusion coefficient and localization are only logarithmically frequency dependence. Increasing the frequency over $n$ orders of magnitude, we observe that the value $\alpha$ at the localization transition is reduced by a factor of $n$. In contrast to $d_H=2$, this suggests that the localizing effects of random geometry are enhanced at high frequency (though exponentially small).

For $d_H=0$, we find that transport is completely determined by the short-distance cutoff of the theory, where the continuum description breaks down. Upon further analysis, this failure can be traced back to the unphysical nature of $d_H=0$ membranes. Since there is no correlation in amplitude of the quenched height field between arbitrarily close points in space, derivatives can become arbitrarily large as the lattice spacing goes to zero, resulting in a lack of a well-defined curvature. We can still analyze the theory, however, and we find the diffusion coefficient to be $\sim \omega^{-1}$. This supports the claim that $d_H=1$ is the marginal case; for short range disorder ($d_H<1$) geometry acs as a high-pass filter, and for longer range disorder ($d_H>1$) it acts as a low-pass filter.

The unphysical dependence on $\Lambda$ plaguing the $d_H=0$ case, also appears to an extent for $d_H=1$. An alternative way to express the claim that $d_H=1$ represents the marginal case, is by looking at its $\Lambda$ dependence. Membranes belonging to the $d_H=0$ ensemble exhibit are $\sim \Lambda^2$, in the $d_H=1$ ensemble $\sim \ln\Lambda$, and in the $d_H=2$ ensemble $\Lambda$-independent.

In all cases of disorder, the localization length is found to depend exponentially on $\alpha$ and $\omega$. This is a feature of two dimensional systems, and indicates that the divide between localized/extended states is sharp.

In future work we would like to explore fluctuation corrections to our results. In particular, it would be interesting to intensity fluctuations in the diffusive limit, and see if the system obeys a type of geometrical speckle-correlation. Additionally, we would like to understand the sensitivity of our results to inelastic scattering. In biological applications, membranes are immersed in viscous fluid. Whether or not localization effects persist in the overdamped limit is a question of interest.

\section{acknowledgements}
The authors would like to thank Valentin Slepukhin for thoughtful discussions. This work was partially supported by DMR grant 000-4566 


\appendix

\section{Self-energy calculation}
\label{app: self-energy}
The reader primarily interested in the results, is encouraged to skip directly to Tab.~\ref{tab: self energy data}.

\begin{figure}
\includegraphics[scale=0.7]{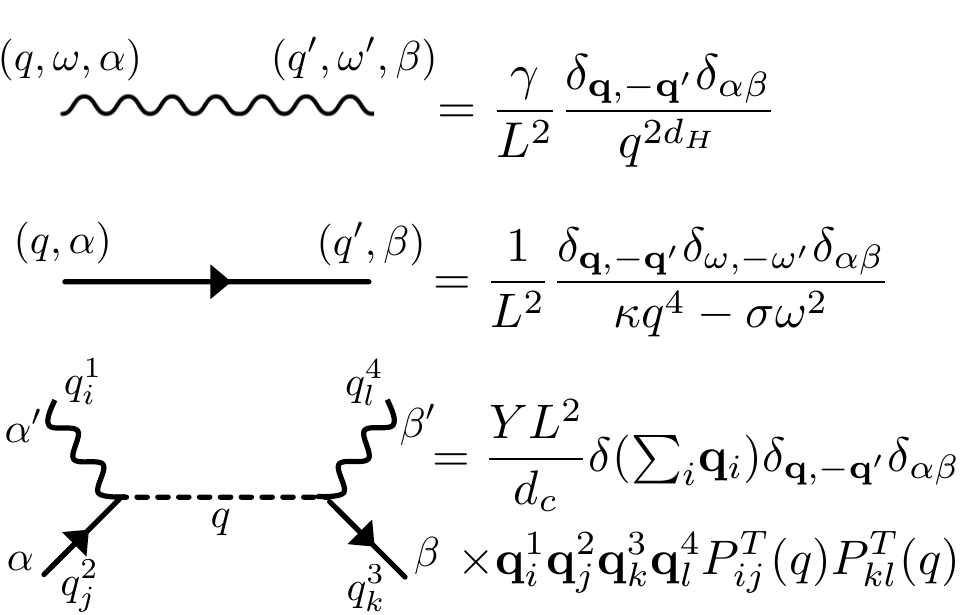}
\caption{Straight (wavy) lines represent propagators for the undulation (disorder) fields. The effective disorder vertex corresponding to Eq.~\ref{eq: Sint}, is on the third line. The vertex carries factors of wavenumber that can be accounted for by the following rule: each line (both wavy and solid) carries one factor wavenumber for each intersection that it terminates at. Only diagrams that remain connected when disorder lines are cut contribute to disorder averages.}
\label{fig: diagrampieces}
\end{figure}

We compute the disorder averaged Green's function $\langle G^+ \rangle$ and thereby, via Eq.~\ref{eq: G def}, the self-energy. The field theory is defined by the action in Eq.~\ref{eq: S}, and the perturbation theory by the subsequent decomposition of $S$ into a Gaussian piece $S_0$, and an interacting piece $S_\text{int}$. The dimensionless parameter regulating the perturbation series is determined post factum after computing the first order correction. The elementary propagators and vertices are shown diagrammatically in~Fig.~\ref{fig: diagrampieces}.
All calculations are performed using the large $L$ limit, whereby we replace summations $L^{-2}\sum_p (\ldots)$ with integrations $\int \frac{d^2 p}{(2\pi)^2}(\ldots)$. The projection operators appearing in the disorder vertex can alternatively be written as the cross product of two two-dimensional vectors projected in to three dimensions as
\beq
\label{eq: cross prod}
p_i P_{ij}^T(q) p_j =  \frac{|p \times q|^2}{q^2},
\eeq
which will prove useful. Since $p$ and $q$ are not actually vectors, we omit the bold face, with the understanding that cross products are evaluated according to Eq.~\ref{eq: cross prod}

\begin{figure}
\includegraphics[scale=.65]{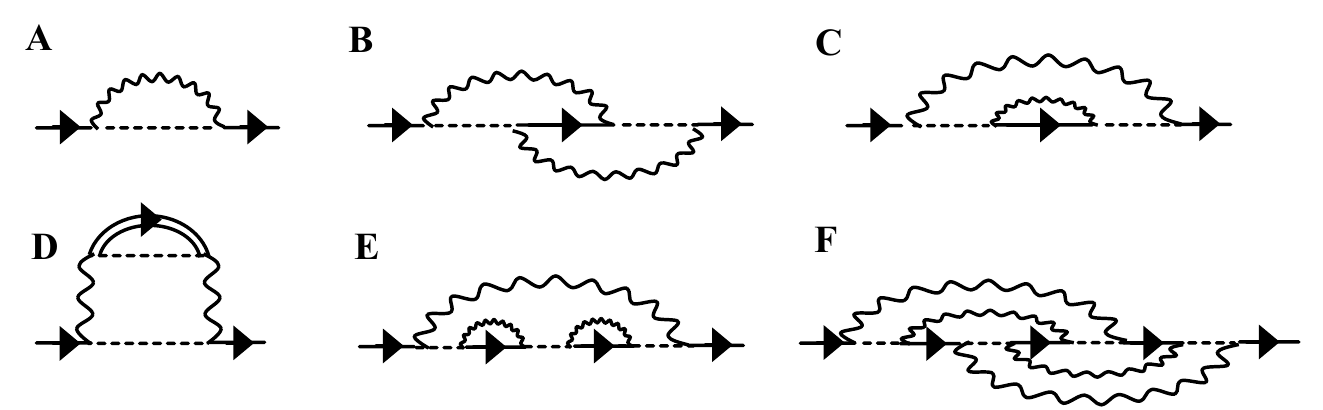}
\caption{A: The first order correction to the self-energy and Green's function. The absence of an internal propagator makes this contribution purely real. B: Second order correction of higher order $\mathcal{O}(1/d_c^{2})$. C: Second order term in SCSA and first order term for the imaginary component of self energy. D: An example of forbidden diagrams, in which a horizontal cut across disorder lines leaves the graph disconnected. Double solid lines indicate fully dressed Green's function. E: Third order term in SCSA, but left out by the NCA. F: An example of a crossed diagram, whose phase space is restricted, leading to a result of higher order in $1/(p_F \ell)$. These diagrams are small in both the SCSA and NCA approximations.}
\label{fig: examplediagrams}
\end{figure}

The rule for contracting lines is slightly different than for typical field theories due to the nature of the disorder average. In any diagram, one must first contract all solid lines in order to build the propagators $\hat G^+,\, \hat G^-$, then afterwards contract the remaining wavy lines to perform the disorder average. This is accounted for by implementing the additional rule that only diagrams that remain fully connected when all disorder lines are cut may contribute to any given calculation. An example of a particular class of forbidden diagrams is shown in Fig.~\ref{fig: examplediagrams}D. 

As usual, the self energy is given by the set of one-particle irreducible diagrams, {\em i.e.}~diagrams that remain connected after an solid line is removed. Per the disorder rules, dashed lines and solid lines count toward connectivity, but wavy lines do not.

\subsection{perturbation series}
\label{app: perturbation series}
The lowest order term for the self-energy, shown in Fig.~\ref{fig: examplediagrams}A, is equal to the equation
\beq
\label{sigma0 integral}
\Sigma^{(1)}_p = \frac{-\gamma Y L^2}{d_c} \int \frac{d^2q}{(2\pi)^2} \frac{ |p \times q|^4}{q^4(q-p)^{2 d_H}}.
\eeq
Since there is no imaginary part or $\omega$ dependence, there is no distinction between advanced/retarded and we thus omit the $\pm$ index. Counting powers of $q$ we see that the integrand $\sim q^{2-2d_H}$, which indicates a divergence at high (low) wavenumber for $d_H$ less (greater) than one. We regulate the high wavenumber divergence by imposing an upper cutoff $\Lambda$ beyond which the continuum model breaks down. 
If the membrane possess an underlying lattice structure, $\Lambda$ is on the order of the inverse lattice spacing (or grid spacing in numerical calculations). Integrating the angular components we find
\beq
\Sigma^{(1)}_p = \frac{-\gamma Y L^2 p^{6-2 d_H}}{4\pi^2 d_c} \int_0^{\Lambda/p} q I_{d_H}(q) dq,
\eeq
where we have defined the commonly occurring function
\beq
I_{d_H}(p)= \int_0^{2\pi} \frac{\sin^4 \theta \, d\theta}{(p^2 + 1 - 2 p \cos \theta)^{d_H}}.
\eeq
$I_{d_H}$ is easily solved by substitution $z = e^{i \theta}$, followed by a contour integration around the unit circle. The results for $d_H=0,1,2$ are summarized in Tab.~\ref{tab: Idh}.

\begin{table}
\caption{
Values of $I_{d_H}(q)$ for $d_H=0,1,2$.}
\label{tab: Idh}
\begin{ruledtabular}
\begin{tabular}{cccc}
 $$
 &$ d_H=0 $
  &$ d_H=1 $
   &$ d_H=2$
\\ \hline
$I_{d_H}(q)$
 &$\frac{3 \pi}{4}$
 &$\frac{3\pi}{4 \max(q,1)^2}\left(1-\frac{\min(q,1)^2}{3 \max(q,1)^2}\right)$
 &$\frac{3 \pi}{4 \max(q,1)^2}$
\end{tabular}
\end{ruledtabular}
\end{table}

Substituting and performing the radial integration we find
\beq
\Sigma_p^{(1)} = \frac{-3 \gamma Y L^2 p^4}{16 \pi d_c} \times \left \{
\begin{array}{cc}
\Lambda^2/2  & d_H =0 \\
\frac{p^2}{6 \Lambda ^2}+\ln \left(\frac{\Lambda }{p}\right)+\frac{1}{4}   & d_H=1 \\
\frac{1}{p^2}-\frac{1}{2 \Lambda ^2}  & d_H =2
\end{array}.
\right.
\eeq
The $\Lambda$ divergence is due to the lack of a well-defined curvature tensor for surfaces $d_H<2$, which can be inferred from the large $p$ limit of $\langle |\nabla^2 h| \rangle \sim p^{2-d_H}$. This suggests that the weak scattering approximation is only realizable (and physically meaningful) for $d_H \geq 1$. 

We infer that the perturbation series is regulated by the $d_H$-dependent parameter~\cite{kovsmrlj2013mechanical}
\beq
\frac{Y h_{\text{eff}}^2}{\kappa} <1,
\eeq
where $h_\text{eff}$ denotes an effective averaged height field
\beq
\label{eq: heff}
h_{\text{eff}}^2(q) \sim \left \{ \begin{array}{cc} q^{2-2d_H} & d_H \geq 2 \\ \ln \Lambda/q  & d_H = 1 \\ \Lambda^{2-2d_H} & d_H<1 \end{array} \right. .
\eeq

The strong dependence on $d_H$ has dramatic consequences for the effective elastic constants of warped membranes, leading to a system size dependent rigidity $\kappa \sim L$ for $d_H=2$, compared to only a weak logarithmic $\kappa \sim \ln L$ and system size independent scaling for $d_H=1$ and $d_H =0$ respectively~\cite{kovsmrlj2013mechanical}.

Keeping only the lowest order contribution to the self energy is plagued by two issues. The first is the dependence on the of short distance cutoff $\Lambda$, which causes the perturbation series to diverge. The second, and more important, is the lack of an imaginary component, which is necessary to describe scattering. The lowest order contribution to $\text{Im}\Sigma$ occurs at two loop order (see Figs.~\ref{fig: examplediagrams}B and~\ref{fig: examplediagrams}C).

Both of these problems are treated by performing a partial resummation of the perturbation series known as the self-consistent screening approximation (SCSA)~\cite{radzihovsky1992scsa,kovsmrlj2013mechanical,kovsmrlj2014thermal}. We now turn to a calculation of the SCSA self energy

\subsection{self-consistent screening approximation}
The SCSA has proven successful in describing the thermal fluctuations of warped membranes~\cite{radzihovsky1992scsa,kovsmrlj2013mechanical,kovsmrlj2014thermal}. It is exact in the limit $d_c \to \infty$, and corresponds to the re-summation of all diagrams at $\mathcal{O}(d_c^{-1})$. In the example diagrams shown in Fig.~\ref{fig: examplediagrams}, C and E are $\mathcal{O}(d_c^{-1})$ and contribute to the SCSA, while B and F are $\mathcal{O}(d_c^{-2})$ and $\mathcal{O}(d_c^{-4})$ respectively, and do not. The latter two admit crossed disorder lines. The SCSA can be viewed as a generalization of the non-crossing approximation used in electron transport calculations~\cite{rammer2018quantum}.

\begin{figure}
\includegraphics[scale=0.8]{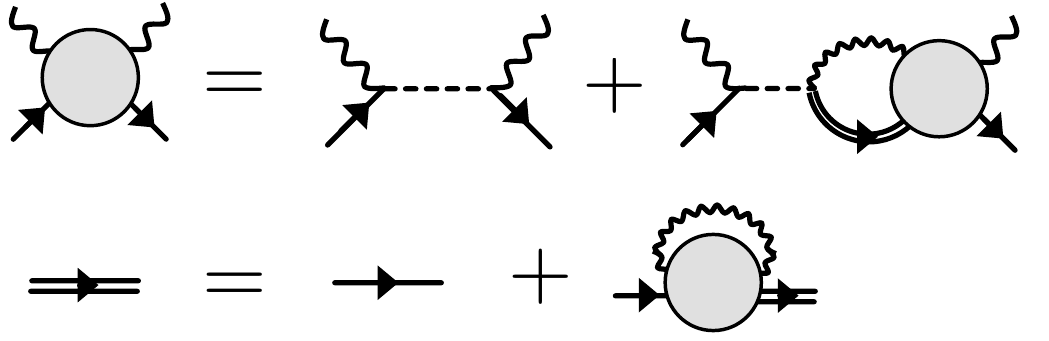}
\caption{The full disorder averaged propagator is represented by a double solid line, and the full vertex by a shaded bubble with two wavy lines and two solid lines attached. Top: The SCSA for the fully dressed Green's function. Bottom: The SCSA for the renormalized vertex function}
\label{fig: SCSA}
\end{figure}

The resummation of all $\mathcal{O}(d_c^{-1})$ diagrams is done diagrammatically in Fig.~\ref{fig: SCSA}. This is equivalent to the set of self-consistent equations

\begin{subequations}
\beq \begin{aligned} 
\label{eq: SCSA G}
G_p^\pm &= L^{-2}(\kappa p^4 - \sigma \omega^2 \mp i \epsilon)^{-1} - \\ 
&\frac{2 d_c^{-1} L^{-2}}{\kappa p4 - \sigma \omega^2 \mp i \epsilon} \sum_q \frac{p_i p_j R^\text{SCSA}_{ij,kl}(q) p_k p_l}{(q-p)^{2 d_H}(\kappa p^4 - \sigma \omega^2 - L^{-2}\Sigma_p^\pm) },
\end{aligned} 
\eeq
\beq 
\label{eq: SCSA R}
\begin{aligned} R^\text{SCSA}_{ij,kl}(q) = R_{ij,kl}(q)- \sum_p\frac{ 2L^{2}R_{ij,mn}(q) p_m p_n p_r p_s R^\text{SCSA}_{rs,kl}(q)}{(p-q)^{2 d_H} (\kappa p^4 - \sigma \omega^2 - \Sigma_p^\pm)},
\end{aligned} 
\eeq
\end{subequations}
describing the dressed propagator and vertex.

We simplify these equations as follows. We first rewrite Eq.~\ref{eq: SCSA G} in terms of the self energy by multiplying both sides by $L^{2}(\kappa p^4 - \sigma \omega^2 \mp i \epsilon)(G_p^\pm)^{-1}$ and using the definition in Eq.~\ref{eq: G def}. Rearranging leads to
\beq
\label{eq: sigma scsa}
\Sigma_p^\pm = \frac{-2}{d_c L^2}\sum_p\frac{ p_i p_j R^\text{SCSA}_{ij,kl}(q) p_k p_l}{(p-q)^{2 d_H}}.
\eeq
The tensor indices appearing in $R^\text{SCSA}_{ij,kl}$ are removed by assuming a solution of the form
\beq
\label{eq: R ansatz}
R^\text{SCSA}_{ij,kl}(q) = Y_R(q) P_{ij}^T(q) P^T_{kl}(q),
\eeq
which amounts to a renormalization of the Young's modulus $Y$. Insertion into Eq.~\ref{eq: SCSA R} immediately yields the solution
\beq
Y_R(q) = \frac{Y}{1+\Pi_q},
\eeq
where we have defined the function
\beq
\Pi_q = \frac{\gamma Y}{L^2} \sum_{q'} \frac{\left[q'_i P_{ij}^T(q) q'_j\right]^2}{(q-q')^{2d_H}(\kappa q'^4-\sigma \omega^2-\Sigma_{q'} L^{-2})}.
\eeq
Inputting Eq.~\ref{eq: R ansatz} into Eq.~\ref{eq: sigma scsa} we complete our setup of the SCSA. This has been reduced to solving the set of self consistent equations

\begin{subequations}
\beq
\label{sigma SCSA def}
\Sigma_p = \frac{-\gamma Y}{d_c} \sum_q  \frac{ |p \times q|^4}{q^4(q-p)^{2d_H}(1+\Pi_q)},
\eeq
\beq
\label{Pi SCSA def}
\Pi_q = \frac{\gamma Y}{L^2} \sum_{q'} \frac{|q \times q'|^4}{q^4(q-q')^{2d_H}(\kappa q'^4-\sigma\omega^2-L^{-2}\Sigma_{q'} )},
\eeq
\end{subequations}
where we have made use of Eq.~\ref{eq: cross prod}. These equations must be solved for each of the cases $d_H=0,1,2$.

For the remainder of the section, $\omega,p$ refer to their dimensionless versions defined in Sec.~\ref{sec: results}, Eq.~\ref{eq: dimless}. We further work with the dimensionless self-energy $\overline{\Sigma}_p$ defined in Eq.~\ref{eq: dimsig} and the dimensionless disorder amplitude $\alpha$ defined in Eq.~\ref{eq: alpha}. The upper cutoff has units of inverse length and is also nondimensionlized. The dimensionless SCSA equations are now
\begin{subequations}
\label{eq: SCSA}
\begin{align}
\Pi_p &=16 \pi \alpha \int\frac{d^2q}{(2\pi)^2}
 \frac{|p\times q|^4}{p^4(p-q)^{2 d_H}( q^4 - \omega^2 - \overline{\Sigma}_{q} )} 
 \label{eq: SCSA pi}
 \\
\overline{\Sigma}_p &= \frac{-16 \pi \alpha}{d_c} \int\frac{d^2q}{(2\pi)^2} \frac{|p \times q|^4}{q^4(p-q)^{2d_H}(1+\Pi_{q})} .
\label{eq: SCSA sigma}
\end{align}
\end{subequations}
We can further perform the angular integrations, to arrive at 

\begin{subequations}
\label{eq: SCSA numerical}
\begin{align}
\Pi_p &=\frac{4 \alpha}{\pi} \int_0^{\Lambda/p}
 \frac{q^{5-2d_H} I_{d_H}(q)\, dq}{q^4 - \omega^2 - \overline{\Sigma}_{q} } 
 \label{eq: SCSA pi numerical}
 \\
\overline{\Sigma}_p &= \frac{-4 \alpha p^{4-2d_H}}{ \pi d_c} \int_0^{\Lambda/p} \frac{ q I_{d_H}(q) \, dq}{1+\Pi_{q}} .
\label{eq: SCSA sigma numerical}
\end{align}
\end{subequations}
This form is suited for numerical evaluation, and is used to provide a check on our analytical solutions. The imaginary part of the self energy
\beq
\text{Im}\overline{\Sigma}_p =\frac{-4 \alpha p^{4-2d_H}}{\pi d_c} \int_0^{\Lambda/p} \frac{q I_{d_H}(q) \text{Im}\Pi_q}{|1+\Pi_q|^2} dq,
\eeq
is $\Pi$-dependent. In the weak scattering limit, we expect $Y_R(p_F)$ is not significantly renormalized, which implies that $\Pi(p_F)$ is small. We thus approximate $\text{Im}(1+\Pi)^{-1} \approx \text{Im} \Pi$. This approximation ignores vertex renormalization and is equivalent to the self-consistent diagrammatic equation
\beq
\label{eq: nca}
\text{Im}\overline{\Sigma}_p \approx \text{Im}\,\includegraphics[width=0.35\linewidth, valign=c]{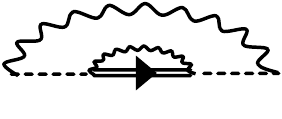}.
\eeq

In the weak scattering approximation, $\text{Im}\Pi(q)$ can be simplified using Eq.~\ref{eq: delta G approximation} to eliminate the radial $q'$ integration. We focus on the particular solution when $p=p_F$ which is relevant for the scattering time and mean free path. We obtain
\beq
\text{Im}\overline{\Sigma}_{p_F} = \frac{-4 \alpha^2 p_F^{8-4d_H}}{\pi d_c (1-\delta_1)}\int_0^{\Lambda/p_F} q I_{d_H}^2(q) dq.
\eeq
Explicitly for $d_H=0,1,2$:
\beq
\label{eq: im sig}
\text{Im}\overline{\Sigma}_{p_F} = \frac{\pi \alpha^2}{d_c (1-\delta_1)} 
\times
\left\{ \begin{array}{cc} 
 9\Lambda^2 p_F^6/8 & d_H = 0 \\
 19 p_F^4/12 & d_H=1 \\
3/2 & d_H =2
\end{array}\right.
\eeq

The fermi wavenumber $p_F$ and $\delta_1$ both depend on the real part of the self-energy, and hence must be checked to be consistent with the SCSA. 

In the following sections we solve the SCSA equations to determine the self energy. The main results are collected in Tab.~\ref{tab: self energy data}.

\begin{table*}
\caption{We tabulate the main results of the SCSA. The imaginary part of the self energy is given in the weak scattering limit.}
\begin{ruledtabular}
\begin{tabular}{l c c c}
 $d_H$
 	&$p_F$
 	&$\text{Re}\overline{\Sigma}_p$
 	&$\text{Im}\overline{\Sigma}_{p_F}$
\\ \hline
0
 	&$\omega^{1/2} (3 \alpha \Lambda^2/2)^{-1/6}$
	 &$-(3\alpha \omega/2)^{1/2} \Lambda p^3$
 	&$\frac{9 \pi \alpha^2 \Lambda^2 p_F^6 /8}{1+\frac{1}{2}(3/2)^{3/2} \Lambda \alpha^{1/2}\omega^{1/2} /p_F^{2}}$
 \\
 1
 	&$ \frac{2^{1/4} \sqrt{\omega}}{\left(1 + \sqrt{1+12 \alpha (\ln(\Lambda/\sqrt{\omega}) +1/4)}\right)^{1/4}}$
 	&$\frac{q^4}{2}\left(1-\sqrt{1+12\alpha (\ln(\Lambda/q)+1/4)}\right)$
 	&$\frac{19 \pi \alpha^2 p_F^4/6}{1+\frac{3\alpha/2 +12 \alpha \ln(\Lambda/p_F) +1}{\sqrt{3\alpha +12 \alpha \ln(\Lambda/p_F) +1}}}$
 \\
2
 	&$\sqrt{\omega}$\footnote{in the limit $\alpha \ll \omega$},\, $\frac{\sqrt{\omega }}{\sqrt[4]{\frac{3}{2} \ln \left(\frac{\alpha }{\omega }\right)+\frac{7}{4}}}$\footnote{in the limit $\omega \ll \alpha $}
 	&$\frac{-3 \alpha p^2}{2} \left(1- \frac{\alpha}{p^2} \ln(1+\frac{p^2}{\alpha}) + \frac{p^2}{\alpha}\ln(1+\frac{\alpha}{p^2})\right)$
 	&$\frac{3 \pi \alpha^2/2}{1+\frac{3}{2} \ln (1+\alpha/p_F^2)}$
\end{tabular}
\end{ruledtabular}
\label{tab: self energy data} 
\end{table*}

\subsubsection{iterative solution}
In principle, the SCSA Eqs.~\ref{eq: SCSA sigma numerical},~\ref{eq: SCSA pi numerical} can be solved via iteration. This is done by first guessing zeroth order solutions (we say order, though there is no formal order parameter governing the iteration) $\overline{\Sigma}_p^{(0)},\Pi^{(0)}$, then performing the integrations on the right hand side to obtain new solutions $\overline{\Sigma}_p^{(1)},\Pi_p^{(1)}$. These, in turn, are put into the right hand side and integrated over yielding another set of solutions $\overline{\Sigma}_p^{(2)},\Pi_p^{(2)}$. The process is repeated ad infinitum until a stationary solution is reached, {\em i.e.} the $n^\text{th}$ order solution is equal to the $(n-1)^\text{th}$ to within some desired tolerance. We make use of this method to develop an approximate solution after only a couple of iterations. 

Our first task is to determine a reasonable zeroth order solution. The simplest function we can construct is a power law $\overline{\Sigma}_q^{(0)},\Pi_q^{(0)} \sim q^{\eta_s,\eta_p}$. The solution is $\omega$-independent, and the exponents $\eta_s,\eta_p$ can be determined by power counting. If $0 \leq \eta_s\leq 4$, then Eq.~\ref{eq: SCSA pi} implies $\Pi_q^{(0)}\sim q^{2-2d_H}$, {\em i.e.} $\eta_p = 2-2d_H$. 

For $d_H \geq 1$, $\Pi_q^{(0)}$ is a decreasing function which does not contribute to power counting on the right side of Eq.~\ref{eq: SCSA sigma} at large $q$. Since the integral is peaked around the value ${\bf q}\approx {\bf p}$, the small $q$ divergence of $\Pi_q^{(0)}$ will not significantly alter the integration so long as $p$ is not much less than one. The scaling of $\overline{\Sigma}$ is then $\overline{\Sigma}_q^{(0)} \sim q^{6-2d_H}$, which is indeed self-consistent.

For $d_H<1$, the integration on the right side of Eq.~\ref{eq: SCSA pi} diverges unless $\eta_s >4$. However, since the integration on the right side of Eq.~\ref{eq: SCSA sigma} cannot push $\eta_s$ greater than four, it must be the case that $\Pi_q^{(0)}$ depends on the upper cutoff as $\Pi_q^{(0)} \sim \Lambda^{2-2 d_H}$. Likewise, the integration on the right side of Eq.~\ref{eq: SCSA sigma} also depends on the upper cutoff $\sim \Lambda^{2-2 d_H}$. We then find the set of self-consistent solutions $\overline{\Sigma}_q^{(0)} \sim q^4 \Lambda^{2-2d_H}$ and $\Pi_q^{(0)}\sim \Lambda^{2-2d_H}$.

\begin{table}
\caption{The zeroth order trial solutions in the iterative method of solving the SCSA equations (Eq.~\ref{eq: SCSA}). Imaginary parts appear at first order in the iterative solution.}
\label{tab: zero order}
\begin{ruledtabular}
\begin{tabular}{lll}
$d_H$&$\overline{\Sigma}_q^{(0)}$&$\Pi_q^{(0)}$\\
\hline
0 	& $\sim q^4 \Lambda^2$ 	& $\sim \Lambda^2$    \\
1 	& $\sim q^4 $			& $\sim \ln \Lambda/q$  \\
2   	& $\sim q^2$			& $\sim q^{-2}$    \\\end{tabular}
\end{ruledtabular}
\end{table}

The zeroth order solutions for each case of $d_H$ are summarized in Tab.~\ref{tab: zero order}. Next we insert these into Eq.~\ref{eq: SCSA} and perform the integrations to obtain the first order solutions $\overline{\Sigma}_q^{(1)},\Pi_q^{(1)}$, at which point we terminate the iteration. This step is performed individually for each of the three cases $d_H=0,1,2$. Finally, the first order solutions are inserted into Eq.~\ref{eq: SCSA numerical} numerically evaluated to assess how closely they satisfy the SCSA.

\subsubsection{$d_h =2$}
Per Tab.~\ref{tab: zero order}, we use $\Pi_q^{(0)}\sim q^{-2}$ and $\overline{\Sigma}_q^{(0)} \sim q^2$. Since $\Sigma_q^{(0)}$ does not scale with a power higher than $q^4$, we expect the evaluation of the integrand in Eq.~\ref{eq: SCSA pi} to depend only weakly on $\overline{\Sigma}_q^{(0)}$,and thus solve for $\Pi_q^{(1)}$ by setting $\overline{\Sigma}_q^{(0)}=0$. The second term in the denominator of Eq.~\ref{eq: SCSA pi} is a completed square, which we factorize into
\beq
\begin{split}
\Pi_q^{(1)} =\frac{16 \pi \alpha}{p^4} \int \frac{d^2q}{4\pi^2} \frac{p_a p_b p_c p_d \epsilon_{ai}\epsilon_{bj}\epsilon_{cm}\epsilon_{dn} q_i q_j q_m q_n}{(q-p)^4 (q^2 - \omega \mp i \epsilon)(q^2 + \omega \pm i \epsilon))}.
\end{split}
\eeq
Here, $\epsilon_{ij}$ represents the two-dimensional antisymmetric Levi-Civita symbol, and $i\epsilon$ an infinitesimal imaginary parameter taken to zero at the end of the calculation that keeps track of the retarded/advanced distinction. The remaining integral can now be performed via the method of Feynman parameters~\cite{schwartz2014quantum}. We obtain

\beq
\label{eq:SCSA pi dh2}
\Pi_q^{(1)} = \frac{3 \alpha \omega}{4 q^4}
\left(
i \pi
+ \frac{2q^2}{\omega}
+ \left(\frac{q^4}{\omega^2}-1\right)\ln\left( \frac{q^2+\omega}{q^2-\omega}\right) \right)
\eeq

The zeroth order divergence at small $q$ is now cured at first order, with Re$\Pi_{q\to0}^{(1)} \sim q^2/\omega^2$. For large $q$, the leading term is Re$\Pi_{q\to\infty}^{(1)} = 3/q^2$. This has the same behavior as the zeroth order ansatz, and we shall use this to fix the proportionality constant, {\em i.e.}\ $\Pi_q^{(0)} = 3/q^2$.

The integral in Eq.~\ref{eq: SCSA sigma} is performed using the same techniques as before, with the result

\beq
\label{eq:SCSA real sigma dh2}
\text{Re}  \overline{\Sigma}_q^{(1)} = \frac{-3\alpha q^2}{2 d_c } \left[1 -\frac{\alpha}{q^2}  \ln \left(1+\frac{q^2}{\alpha}\right) +\frac{q^2}{\alpha} \ln \left(1+\frac{\alpha}{q^2}\right)\right]
\eeq

From this, we may solve for renormalized group velocity and Fermi wavenumber (see Eqs.~\ref{eq: group velocity deltas} and~\ref{eq: fermi momentum} respectively). Trivially, $\delta_2=0$ and
\beq
\delta_1=\frac{-3}{2 d_c} \ln \left(1+\frac{\alpha}{q^2}\right).
\eeq

The Fermi wavenumber, $p_F$, is given by the solution to the nonlinear equation $p_F^4 - \omega^2 = \text{Re}\overline{\Sigma}_{p_F}$. We can find a solution in two limits. For $\alpha \ll \omega$, $\text{Re}\overline{\Sigma} = -3 \alpha p^2$ yielding a quadratic equation for $p_F^2$. In this limit, we use
\beq
p_F(\alpha \ll \omega) = \sqrt{\frac{1}{2} \left(\sqrt{9 \alpha ^2+4 \omega ^2}-3 \alpha \right)}.
\eeq
In the opposite limit, we find that $\text{Re}\Sigma_p \to \frac{-3}{4}\left(q^4+2 q^4 \ln \left(\alpha /q^2\right)\right)$. This yields an approximate solution
\beq
p_F(\alpha \gg \omega) = \frac{\sqrt{\omega }}{\sqrt[4]{\frac{7}{4}+\frac{3}{2}\ln \left(\frac{\alpha }{\omega }\right)}}.
\eeq

\subsubsection{$d_H=1$}
Per Tab.~\ref{tab: zero order}, we use zero order solutions 
\beq
\label{eq: dh1 zero}
\Pi_q^{(0)} = (c_{\pi}-1), \; \overline{\Sigma}_q^{(0)} = (1-c_s^4)q^4,
\eeq
for some constants $c_{\pi},c_s$ to be determined. Both integrals of the SCSA are logarithmically divergent and depend on the upper cutoff $\Lambda$. Assuming $\Lambda \gg 1$, we discard terms $O(\Lambda^{-1})$. $\overline{\Sigma}_q^{(1)}$ is evaluated easily from Eq.~\ref{eq: SCSA sigma numerical}, giving

\beq
\label{eq: sig dh1 one}
 \overline{\Sigma}_q^{(1)} = \frac{-3 \alpha q^4}{c_{\pi} d_c}\left(\ln(\Lambda/q) + 1/4\right).
\eeq

$\Pi_q^{(1)}$ is calculated from Eq.~\ref{eq: SCSA pi} using the method of Feynman parameters as was done for the $d_h=2$ case. We find
\beq
\label{pi1 dh1 1}
\Pi_q^{(1)} =
\frac{3\alpha}{c_s^4}
\int_0^1 (1-x_1)\left(\ln\left(\frac{c_s^2 \Lambda^2}{\Delta^2 \omega}\right) - \frac{3}{2} \right)dx_1 dx_2,
\eeq
where we have defined
\beq
\Delta^2 = (1-x_1)\left(\frac{c_s^2 q^2}{\omega} x_1 + (2 x_2 -1)\right).
\eeq
The remaining integrations over Feynman parameters $x_1$ and $x_2$, may be carried out to give
\begin{widetext}

\beq
\label{eq: pi dh1 def}
\Pi_1^{d_H=1}(y)=\frac{\alpha}{4 c_s^4  y^4}\left[  -\ln( y^4-1)-6  y^2 \tanh ^{-1}\left( y^2\right)+ y^4 \left(3\ln\left( \frac{c_s^4 \Lambda^4/\omega^2}{ y^4-1}\right)+5 \right)-2  y^6 \coth ^{-1}\left( y^2\right)+i \pi(1- 6  y^2)\right]
\eeq

\end{widetext}
with $y = c_s q/\omega^{1/2}$. We only use this expression insofar as to perform numerical checks on our calculation, since we are principally interested in the calculation of $\overline{\Sigma}_p$. We can construct an approximation of this function as follows.

First, we note that the integrand of Eq.~\ref{pi1 dh1 1} is most strongly peaked at $q=0$ and at $q=\Lambda$. In the former limit, the integrand is determined primarily by the larger of $q^2, \omega$. This suggests that we may approximate $\Pi_1(q)$ as a piecewise function transitioning from the low $q$ to high $q$ behavior near $q \sim \omega^{1/2}$. Specifically, Taylor expanding Eq.~\ref{pi1 dh1 1} at low/high $q$ then solving for the value $q^*$ at which the difference between the two solutions is minimized, we find a transition point $q^*=6^{1/4} \omega^{1/2}$. This yields the approximate solution

\beq
\label{eq: pi dh1 approx}
\text{Re}\Pi_q^{(1)} \approx \frac{3\alpha}{c_s^4}
\left\{
\begin{array}{lc}
\ln \left(c_s \Lambda/\omega^{1/2}\right)-\frac{c_s^4 q^4}{24\omega^2} & q< 6^{1/4}\omega^{1/2}/c_s \\
\ln(\Lambda/q ) +1/4 & q \geq 6^{1/4}\omega^{1/2}/c_s
\end{array}
\right.
\eeq

The constants $c_{\pi},c_s$ are determined by matching the first order solution to the zeroth order solution, which is most easily accomplished in limit $q \to \Lambda$. Critically, our power law analysis of the zeroth order solution omitted non-analytic functions. Eq.~\ref{eq: pi dh1 approx} suggests $c_{\pi},c_s$ are not strictly constant, but can admit logarithmic dependence on $q$. Matching the $q \to \Lambda$ limit of Eqs.~\ref{eq: pi dh1 approx} and~\ref{eq: sig dh1 one} to Eq.~\ref{eq: dh1 zero} yields the set of equations
\begin{subequations}
\beq
c_{\pi}-1 = \frac{ 3 \alpha }{d_c c_s^4} \left( \ln(\Lambda/q)+1/4\right), 
\eeq
\beq
1- c_s^4 =\frac{ -3 \alpha}{d_c c_p} \left( \ln(\Lambda/q)+1/4\right).
\eeq
\end{subequations}
These have the solution
\beq
c_s^4 = \frac{1}{2}\left( 1 \pm \sqrt{1 + 12 \alpha d_c^{-1} \left( \ln(\Lambda/q)+1/4\right)}\right).
\eeq
The condition $c_s^4=1$ at $\alpha=0$ requires that we choose the (+) solution. Finally, we can quickly find $\overline{\Sigma}_q^{(1)}$ by using the matching condition $\overline{\Sigma}_q^{(1)}=(1-c_s^4) q^4$ to find
\beq
\text{Re}\overline{\Sigma}_q^{(1)}= 
\frac{q^4}{2}\left( 1 - \sqrt{1 + \frac{12 \alpha}{d_c} \left(\ln(\Lambda/q)+1/4\right)}\right).
\eeq

We solve for the Fermi wavenumber by evaluating Re$\overline{\Sigma}_p$ at $p=\omega^{1/2}$. This approximation is increasingly accurate as either $\alpha \to0$ and/or $\Lambda \to \infty$. We obtain
\beq
p_F = 2^{1/4}\omega^{1/2}\left( 1 + \sqrt{1 + \frac{12 \alpha}{d_c} \left(\ln(\Lambda/\omega^{1/2})+1/4\right)}\right)^{-1/4}.
\eeq
The group velocity constant $\delta_2=0$, and 
\beq
\delta_1 =\frac{1}{2}-\frac{3\alpha/2 +12 \alpha  \ln \left(\frac{\Lambda }{\omega^{1/2}}\right)+1}{2\sqrt{3 \alpha +12 \alpha  \ln \left(\frac{\Lambda }{\omega^{1/2}}\right)+1}}.
\eeq
In the numerical calculations of Fig.~\ref{fig: scsa numchecks}, the obtained value for $c_s$ is input back into Eq.~\ref{eq: pi dh1 def} to determine $\Pi_q^{(1)}$.

\subsubsection{$d_H=0$}
We start with Eqs.~\ref{eq: SCSA pi numerical} and~\ref{eq: SCSA sigma numerical}. The zero order solutions are
\beq
\Pi_q^{(0)} = c_{\pi}, \;  \overline{\Sigma}_q^{(0)} = c_s q^4,
\eeq
as for $d_H=1$.  The integrations are quadratically divergent, and dominated by the upper wavenumber cutoff. To lowest order in $\Lambda^{-1}$ we find the SCSA equations
\begin{subequations}
\beq
c_{\pi} = \frac{3 \alpha \Lambda^2}{2}\frac{1}{1 - c_s},
\eeq
\beq
c_s = \frac{-3 \alpha \Lambda^2}{2}\frac{1}{d_c(1+c_p)}.
\eeq
\end{subequations}
These have the solution
\beq
\begin{aligned}
c_s^\pm &= \frac{1}{2 d_c} \left(1-\frac{3 \Lambda^2 \alpha \omega}{2 q^2}(1-d_c) \pm \right. \\ 
&\left. \sqrt{ \left(1-\frac{3 \Lambda^2 \alpha \omega}{2 q^2} (1-d_c)\right)^2 + \frac{6 \Lambda^2 \alpha \omega}{q^2} d_c} \right).
\end{aligned}
\eeq
In order to choose the correct branch, we consider the limit $\alpha \to 0$. This corresponds to zero disorder, {\em i.e.} $\gamma \to 0$. In this limit, the self energy should vanish and so $c_s \to 0$. This uniquely singles out the (-) solution $c_s^-$. 

Now we consider the limit $\Lambda \gg 1$. For physical membranes we also set $d_c=1$, which we do first, noting that the limits $d_c\to 1$ and $\Lambda \to \infty$ do not commute. To leading order in $\Lambda$ we find
\beq
c_s  = \left \{ \begin{array}{lr} \frac{- \Lambda}{q}\sqrt{\frac{3 \alpha \omega}{2}}& : d_c=1 \\ \frac{1}{2 d_c} & : d_c \neq 1 \end{array} \right.
\eeq
$c_s$ is a renormalization of the bending rigidity $\kappa \to \kappa(1 + c_s)$. Since by definition $\Lambda >q$ for all $q$, $c_s$ dominates the effective bending rigidity for all but very small disorder and very low frequency. Assuming that $q \ll \Lambda$ (which is consistent with linearized shallow shell theory), the renormalized dimensionless propagator is
\beq
(G_q^\pm)^{-1} \sim \frac{1}{ \sqrt{3\alpha \omega/2} \Lambda q^3 - \omega^2 - i \text{Im}\Sigma_q^\pm}.
\eeq
We easily determine the Fermi wavenumber
\beq 
\label{eq: pf dh0}
p_F=\omega^{1/2} \left(\frac{2}{3 \alpha \Lambda^2}\right)^{1/6}. 
\eeq
The assumption $p_F \ll \Lambda$ is self consistent, as $p_F$ is dampened by a factor of $\Lambda^{1/3}$. The group velocity can be found directly,
\beq
{\bf v}_G =(12 \alpha)^{1/3} \Lambda^{2/3} {\bf q}.
\eeq
The function
\beq
\delta_1 = \frac{-3}{4} \left(\frac{3\alpha \omega}{2}\right)^{1/2} \frac{\Lambda}{q},
\eeq
in combination with Eq.~\ref{eq: im sig}, gives the intermediate expression for the imaginary part of the self-energy
\beq
\text{Im}\overline{\Sigma}_{p_F} =\frac{9 \pi \alpha^2 \Lambda^2 p_F^6 /8}{1+\frac{3}{4} \frac{\Lambda}{p_F}  \sqrt{\frac{3 \alpha  \omega}{2}}}
\eeq
Inputting Eq.~\ref{eq: pf dh0} for $p_F$ then taking the large $\Lambda$ limit we obtain
\beq
\text{Im}\overline{\Sigma}_{p_F} =\left(2/3\right)^{2/3} \pi \omega^3\alpha^{1/3}\Lambda^{-4/3}.
\eeq

\subsubsection{numerical checks}
\begin{figure}
\includegraphics[scale=0.45]{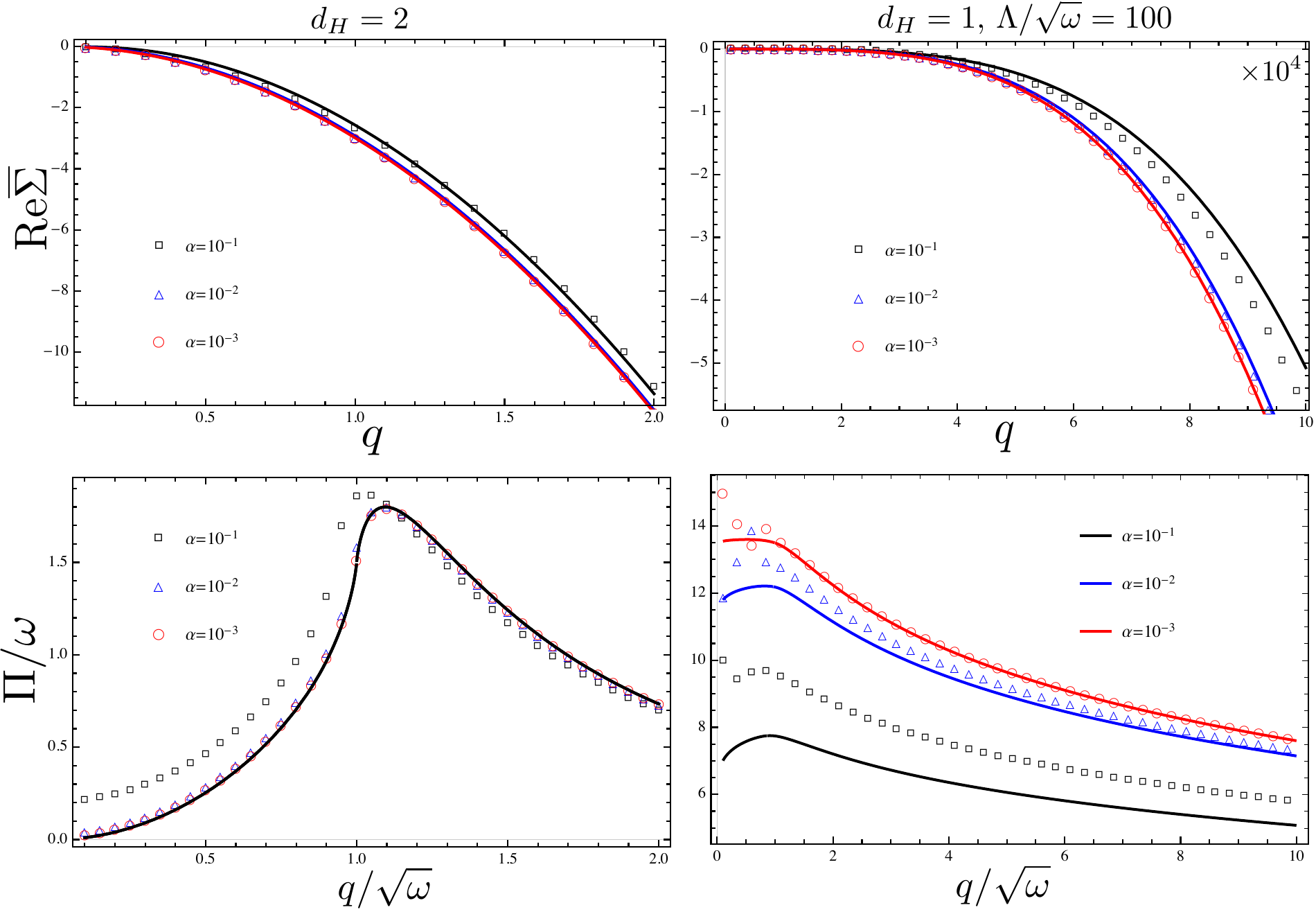}
\caption{Data points represent the numerical integration result for $\text{Re}\overline {\Sigma}_q$ and $\Pi_q$, using the numerical integrations defined in Eqs.~\ref{eq: SCSA pi numerical}~\ref{eq: SCSA sigma numerical}. Curves represent the analytic result. Colors black, blue, and red, and the shapes square, triangle, and circle, represent parameters $\alpha=10^{-1},10^{-2},10^{-3}$ respectively. We consider the cases $d_H=1,2$, where the solutions for $\overline{\Sigma}_q,\Pi_q$ are nontrivial functions. Beyond weak scattering we find our analytic approximations to be inaccurate, however, the shape of the curves is still accurate, which we need for deriving the diffusion coefficient.}
\label{fig: scsa numchecks}
\end{figure}

In Fig.~\ref{fig: scsa numchecks} we numerically test the accuracy of the first order SCSA solutions. This is done by inputting $\overline{\Sigma}_q^{(1)}, \Pi_q^{(1)}$ into Eq.~\ref{eq: SCSA numerical} for the real part of the self-energy, and using Eq.~\ref{eq: im sig} for the imaginary part. The numerical integration is performed at fixed $p$ and compared to the analytic solution. With the exception of $d_H=0$, the self-energy is a $p$-dependent function, so the comparison is done over a range of wavenumbers. For $d_H=0$, we find a single value for $c_s,c_\pi$, in good agreement with the analytical result. For $d_H=1,2$ we find good agreement in the weak scattering approximation for $\alpha =10^{-3}$, with increasing precision for wavenumbers $q>\sqrt{\omega}$ on the order of a percent difference.

\section{$\delta_c$ calculation}
\label{app: delta c}
We begin with the definition of the coherent diffusion coefficient $D_c$ in Eq.~\ref{eq: Dc}. The parameter $\delta_c$ is
\beq
\delta_c = i \tau M_0 .
\eeq
The scattering time $\tau$ was found earlier in Appendix~\ref{app: self-energy} (and tabulated in Tab.~\ref{tab: one particle data}), so we need only evaluate $M_0$, which was defined in Eq.~\ref{eq: M0}. 

\begin{figure}
\includegraphics[scale=0.6]{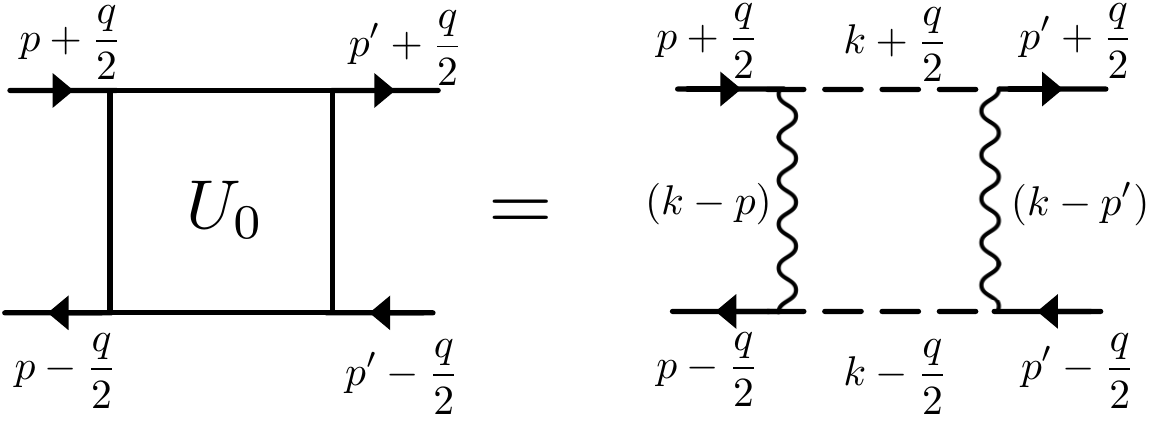}
\caption{The bare irreducible vertex. This is the simplest four-point vertex that can be constructed that remains fully connected after cutting all wavy lines.}
\label{fig: U0}
\end{figure}

\begin{figure}
\includegraphics[scale=.6]{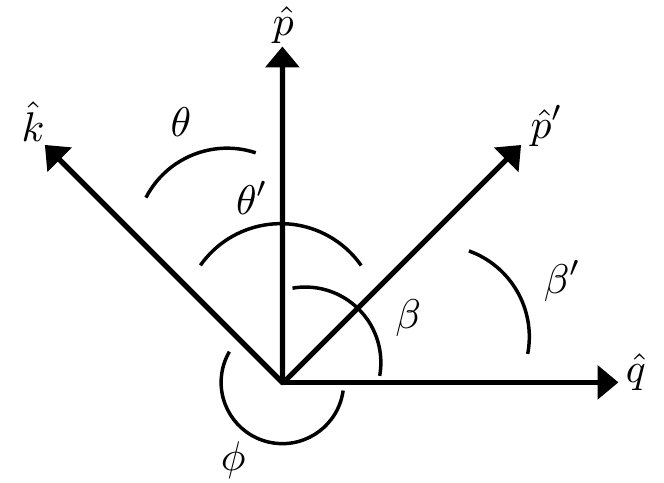}
\caption{Angles between unit vectors at fixed orientation in the calculation of $M_0$. Only $\hat q$ is not integrated over. We first fix $\hat k$, then perform the integrations over $d \hat p = d\theta$ and $d \hat p'= d\theta'$. This yields a function $J_{d_H}(k;\phi)$ that is then integrated over $\phi$.}
\label{fig: angles}
\end{figure}

$D_c$ is found by replacing the irreducible vertex $U$ with the bare vertex $U^0$, defined in Fig.~\ref{fig: U0}. $U^0$ is evaluated at $q=0$ and affords the simpler representation
\beq 
\label{eq: U0}
U^0_{pp'} = \frac{\gamma^2 Y^2}{2d_c} \sum_k \frac{ |p \times k|^4 |k \times p'|^4}{(k-p)^{2d_H}(k-p)^{2d_H}k^8},
\eeq
where the cross product is as defined in Eq.~\ref{eq: cross prod}. We insert this into the definition of $M_0$ then evaluate the corresponding integrals. In the weak scattering approximation, the radial integrations are performed using Eq.~\ref{eq: delta G approximation}, leaving only the angular integrations:
\beq
\label{eq: M0 intermediate}
M_0 = \frac{i\pi \nu}{\sigma^2 \omega^2} \int_0^{2\pi} d \hat p d \hat p' ({\bf \hat q} \cdot {\bf \hat p}) U_{\hat p \hat p'} ({\bf \hat q} \cdot {\bf \hat p'}),
\eeq
and $d \hat p$ is understood to mean the angular integration on $\hat p$. There are four unit vectors to consider, and a total of $4!$ angles to consider. We define angles according to Fig.~\ref{fig: angles}.

The angles $\beta,\beta'$ can be eliminated in favor of $\phi,\theta,\theta'$, allowing us to use ${\bf \hat q} \cdot {\bf \hat p} = \cos(\phi-\theta)$ and ${\bf \hat q} \cdot {\bf \hat p'} = \cos(\phi-\theta')$. Simplifying Eq.~\ref{eq: M0 intermediate} according to Fig.~\ref{fig: angles}, we find
\beq
M_0 = \frac{\pi \nu L^2 p_F^{2-4 d_H}}{2\sigma^2 \omega^2 i d_c} \int_0^{\Lambda/p_F} \frac{kdk}{4\pi^2}\int_0^{2\pi}d\phi J^2_{d_H}(k;\phi),
\eeq
where we have defined
\beq
J_{d_H}(k;\phi) = \int d\theta \frac{\cos (\theta - \phi)\sin^4\theta}{(k^2 + 1 - 2 k \cos \theta)^{2 d_H}}.
\eeq
$J_{d_H}(k;\phi)$ is computed via substitution $z=e^{i\theta}$ followed by contour integration. Performing the remaining integrations yields $M_0$, which is then trivially related to $\delta_c$. The results for both $J_{d_H}(k;\phi)$ and $\delta_c$ are shown in Tab.~\ref{tab: Jdh} for $d_H=0,1,2$. 

\begin{table}
\caption{Angular integration function $J_{d_H}(k;\phi)$ and $\delta_c$ for each of $d_H=0,1,2$. We use dimensionless frequency, wavenumber, and disorder amplitude.}
\begin{ruledtabular}
\begin{tabular}{l c c}
$d_H$ &$\delta_c$  &$J_{d_H}(k;\phi)$\\
\hline
0 	& $0$  &$0$\\
1 	& $\frac{11\alpha^2 p_F^4 \tau}{24\omega}$ & $\frac{\pi \min(k,1) \left(2 \max(k,1)^2 - \min(k,1)^2\right)\cos \phi}{8 \max(k,1)^5}$  \\
2   	&$\frac{3\alpha^2 \tau}{8 \omega}$ & $\frac{\pi \min(k,1) \cos \phi}{2\max(k,1)^5}$  
\end{tabular}
\end{ruledtabular}
\label{tab: Jdh}
\end{table}

In terms of the dimensionless wavenumber, frequency, and disorder amplitude, we find the final results
\beq
D_c = D_0\times \left\{
\begin{array}{cc}
1 & d_H=0 \\
\left(1+\frac{11 \alpha^2 p_F^4 \tau}{24 \omega}\right)^{-1} & d_H=1 \\
\left(1+\frac{3 \alpha^2 \tau}{8 \omega}\right)^{-1}& d_H=2
\end{array}
\right.
\eeq

\section{Ward identity} 
\label{app: WI}
In this section the vector nature of positions and wavenumbers are understood, and we omit bold-face lettering. Additionally, we deviate from the notation of the main text, and use $\hat G^\pm$ to refer to the {\it unaveraged} Green's function. The WI is most easily derived using a functional integral representation for the Green's function~\cite{mckane1981localization}. We define the moment generating function
\beq
Z[j^+,j^-] = \int \mathcal{D} f^+\mathcal{D} f^- e^{- \int A[f^+,f^-]- j^+ f^+ - j^- f^-  d^2x},
\eeq
where we have defined the total action as the sum of retarded/advanced actions
\beq
A[f^+(x),f^-(x)] = A^+[f^+(x)] + A^-[f^-(x)]
\eeq
that, in turn, are decomposed into the sum of a Gaussian action
\beq
A_0^\pm = \frac{\kappa}{2} (\nabla^2 f^\pm)^2 + \frac{\sigma \omega_+^2}{2} (f^\pm)^2,
\eeq
and a disorder interaction
\beq
A_\text{int}^\pm = \frac{Y}{d_c}(\mathcal{\hat L} f^\pm) \nabla^{-4} (\mathcal{\hat L} f^\pm).
\eeq
$A_0^\pm$ and $A_\text{int}^\pm$ are the real space representations of the DMV action in Eq.~\ref{eq: S0} and Eq.~\ref{eq: Sint}. The operator $\mathcal{\hat L}$ was defined in Eq.~\ref{eq: L def}; for our purposes, it is most important to note that it is hermitian, {\em i.e.} for any two functions $g$ and $h$, we have the identity
\beq
\label{eq: L identity}
\int  g \mathcal{\hat L} h \, d^2 x = \int h \mathcal{\hat L} g \,d^2 x.
\eeq
For a fixed realization of disorder, we may obtain the {\it unaveraged} Green's function in the usual way, via functional derivatives:
\beq
G^+(x,x') = \left. \frac{\delta^2}{\delta j^+(x) \delta j^+(x')} \ln Z[j^+, j^-] \right |_{j^\pm=0}.
\eeq
In this section, we will use different notation than the main text, with respect to averaging. The functional integral method first computes the Green's function as the two-point function of $f^+$ with regards to the ensemble dictated by the action. This is done at fixed disorder, and the resulting Green's function must subsequently be averaged over the disorder. We use angular brackets $\langle \ldots \rangle$ to denote averaging over the $f^+$ ensemble, and an overline $\overline{\ldots}$ to denote disorder averaging. In this notation, the Green's function is written as
\beq
\overline{G^+(x,x')} = \overline{\langle f^+(x) f^+(x') \rangle}.
\eeq
The four-point function can similarly be written
\beq
\phi(x,x';y,y') = \overline{ \langle f^+(x)f^+(x') \rangle \langle f^-(y) f^-(y') \rangle}.
\eeq

At zero external frequency $\Omega$, the total action $A$ possesses an $O(2)$ symmetry between retarded/advanced fields, and is invariant under the transformation
\beq
\vv{ f^+ \\ f^-} \to
\left(
\begin{array}{cc}
\cos \theta & \sin \theta \\
-\sin \theta & \cos \theta
\end{array}
\right)
\vv{f^+ \\ f^-}.
\eeq
For nonzero $\Omega$, we perform the change of variables 
\begin{subequations}
\beq
f^+ \to f^+ + \epsilon f^-
\eeq
\beq
f^- \to f^- - \epsilon f^+,
\eeq
\end{subequations}
for $\epsilon$ an infinitesimal parameter. Since $\epsilon$ is small, we Taylor expand the exponential and use invariance of the functional integral change of variables to find the equation
\beq
\langle -2 \sigma \omega \Omega f^+ f^- + j^+ f^- - j^- f^+ \rangle =0.
\eeq
In obtaining this equation, since the realization of disorder is identical for both $f^+$ and $f^-$ fields, the variation $\delta S_\text{int}$ vanishes. Taking two function derivatives $\frac{\delta^2}{\delta j^+ \delta j^-}$, setting $j^+=j^-=0$, then performing the disorder average we find
\beq
2 \omega \Omega \phi(x,x';y,y') = \left[G^+(x,x') - G^-(y,y') \right]\delta(x-y)\delta(x'-y')
\eeq
Using the position space definition of the Green's function
\beq
G^\pm(x,x') = \kappa \nabla^4 - \sigma \omega_\pm^2 - \hat \Sigma^\pm,
\eeq
we can formally divide by $\hat G^+ \hat G^-$ and use the BS equation to find the solution
\beq
\Delta \hat \Sigma^\pm = \hat U \hat \Delta G^\pm.
\eeq
In the wavenumber basis, this takes the simpler form
\beq
\Delta \Sigma_p(q) = \sum_{p'} U_{pp'}(q) \Delta G_{p'}(q).
\eeq
The WI is identical to the well-known result for electrons in disordered media~\cite{vollhardt1992self}. As a check, the WI can easily be seen to hold for the choice of irreducible vertex $U^0_{pp'}(q)$ (Fig.~\ref{fig: U0} and Eq.~\ref{eq: U0}) and self-energy (Eq.~\ref{eq: nca}) used in this manuscript.

\section{Derivation of Diffuson and Cooperon}
\label{app: diffuson}
We begin with deriving the diffuson, which we denote as $\hat \Gamma$. In the position basis, the diffuson is a function of four points $\Gamma(x_1,x_2,x_3,x_4)$, and in the Fourier basis a function of three wavenumbers $\Gamma_{pp'}(q)$ due to translational invariance. The diffuson is an IR divergent four-point vertex, that diverges in the limit $q,\Omega \to 0$. This divergence ensures that even after after disorder averaging, the diffuson is long-range object, and hence represents a two-particle propagator associated with the diffusive dynamics of the intensity field. 

\begin{figure}
\includegraphics[scale =0.6]{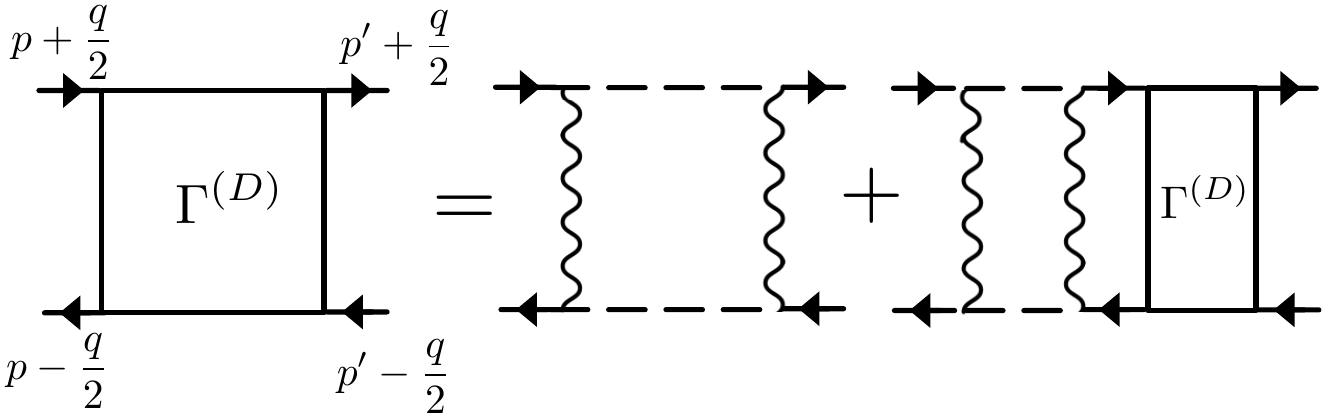}
\includegraphics[scale =0.6]{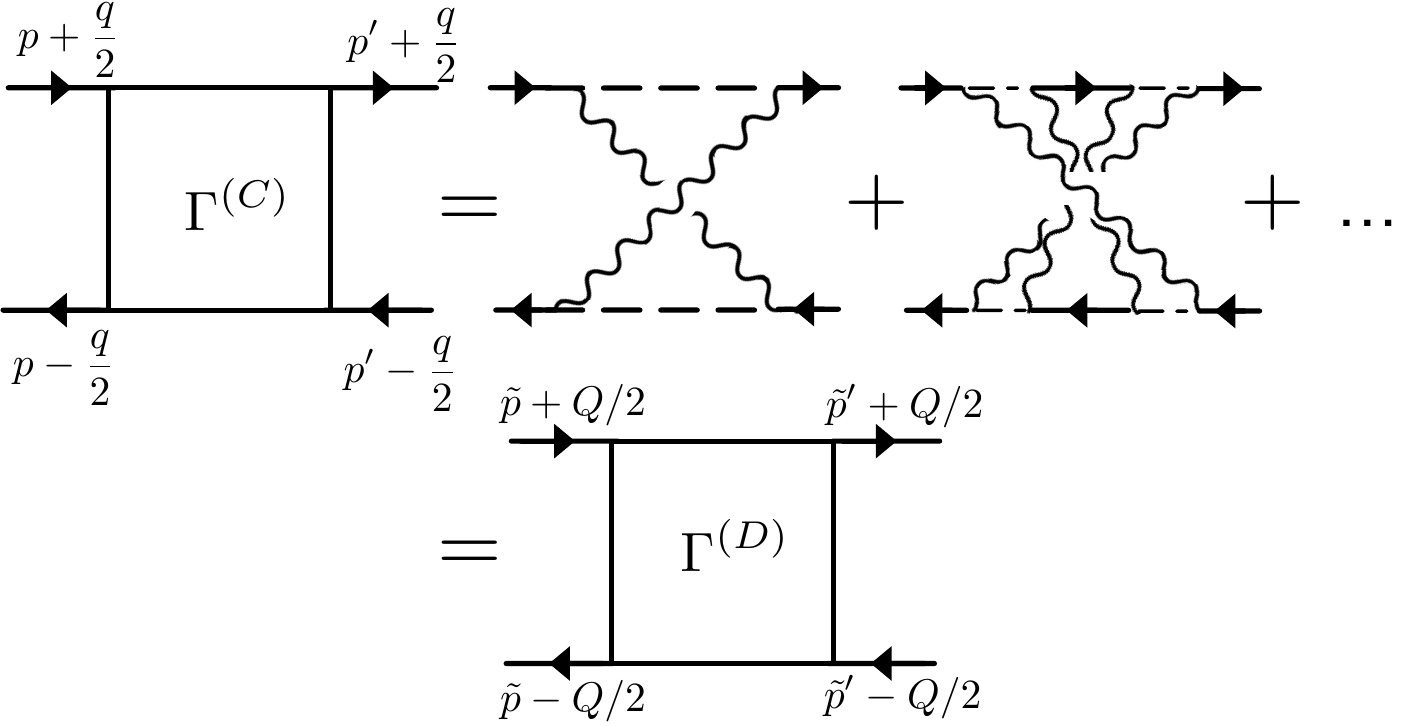}
\caption{Dominant contributions to the four-point vertex in the diffusive limit. Upper/lower lines carry retarded/advanced propagators, each with frequency $\omega +(-) \Omega/2$. The frequency is a passive index and is not integrated over since the disorder field is quenched. Top: diagrammatic representation of the Bethe-Salpeter equation defining the diffuson contribution to the four point vertex $\hat \Gamma$. Bottom: diagrammatic derivation of the cooperon. The new wavenumbers are $\tilde p = \frac{1}{2}(p-p'+q), \,\tilde p' =\frac{1}{2}(p'-p+q),$ and $Q=p+p'$, in agreement with Fig.~\ref{fig: umc}. }
\label{fig: gamma}
\end{figure}

The Green's functions $G^\pm({\bf x}, {\bf x'};\omega)$ represent plane waves of frequency $\omega$ propagating froward/backward (+,-) in time from position ${\bf x'} \to {\bf x}$, and can be interpreted as particles (see section~\ref{sec: localization}). The four-point function is the disorder averaged quantity describing propagation of two paired particles in space. From this representation, we can define the diffuson as the contribution to this amplitude from all paths whereby the paired particles undergo identical scattering paths. In Fourier space, these correspond to the ladder type diagrams of Fig.~\ref{fig: gamma}. 

Summation over ladder diagrams is formally given as a Bethe-Salpeter equation
\beq
\hat \Gamma =\hat  U + \hat U : \hat G^+ \otimes \hat G^- : \hat \Gamma.
\eeq
The colons indicate that $\hat U$ and $\hat \Gamma$ are contracted with the tensor product of $\hat G^+ \otimes \hat G^-$. We keep the same notation as in the text that all Green's functions represent their disorder average. $\hat \Gamma$ decomposes into the two-particle irreducible vertex $U$ (which is currently unspecified), plus the set of ladder diagrams connecting additional vertices by a retarded and advanced propagator. In Fourier space, we find the less abstract form
\beq
\label{eq: gamma BS}
\Gamma_{pp'}(q) = U_{pp'} + \sum_k U_{pk} G_{k+q/2}^+ G_{k-q/2}^- \Gamma_{kp'}(q).
\eeq
Since we are interested in contributions to $\Gamma_{pp'}(q)$ that diverge in the limit $\Omega,q\to0$, we have replaced the first $U_{pp'}(q)$ on the right with its $q=0$ value $U_{pp'}$.

When all of the external legs of the reducible vertex $\hat \Gamma$ are put on shell, we obtain the four-point function of Eq.~\ref{eq: four point}. Explicitly,
\beq
\phi_{pp'}(q) = G_{p+q/2}^+ G_{p-q/2}^- \Gamma_{pp'}(q)G_{p+q/2}^+ G_{p'-q/2}^-.
\eeq
This relation allows us to directly use our results from section~\ref{sec: hydrodynamics} to solve for $\Gamma_{pp'}(q)$. Looking at the small $q,\,\Omega$ limit, we set $q=0$ in the Green's functions and use the identity $G_p^+ G_p^- = \Delta G_p/2i \text{Im}\Sigma_p$. Comparing to the solution for $\phi_{pp'}(q)$ in Eq.~\ref{eq: phi}, we immediately find
\beq
\Gamma_{pp'}^\text{(D)}(q) = \frac{2}{\pi \nu L^2}\frac{\text{Im} \Sigma_p \text{Im}\Sigma_{p'}}{-i \Omega + D_c(\omega) q^2}.
\eeq
The superscript (D) emphasizes this is not the full reducible vertex, but instead its long time/length limit, the {\it diffuson}. The appearance of $D_c$ in the denominator is a consequence of the choice $U=U_0$ for the irreducible vertex, where $U_0$ is the bare irreducible vertex defined in Fig.~\ref{fig: U0}. This choice generates the sum of all non-crossing box diagrams.

The cooperon is derived by first crossing all bare vertices $\hat U^0$, then individually crossing the wavy lines within each $\hat U^0$ as shown in the lower half of Fig.~\ref{fig: gamma}. This is equivalent to a summation over all maximally crossed box diagrams. By left-right reflection of the lower lines ({\em i.e.} reversing all of the arrows), we can un-cross the box diagrams, thereby obtaining an identical set of ladder diagrams as used for the diffuson. The new wavenumbers are reversed and of opposite sign. We thus find the cooperon
\beq
\Gamma^\text{(C)}_{pp'}(q) = \Gamma^\text{(D)}_{\frac{1}{2}(p-p'+q),\frac{1}{2}(p'-p+q)}(p+p').
\eeq
Finally, we note that by including cooperon insertions into the ladder diagrams defining the diffuson~\cite{vollhardt1992self}, we find corrections to the diffusion coefficient $D_c$ that recover the weak localization calculation of section~\ref{sec: localization}.

\begin{widetext}

\section{Full analytic expressions}
\label{app: full expressions}
\subsection{$d_H=0$}
\beq %
D_c = D_0 = \frac{2}{\pi}\frac{(2/3)^{1/3}}{\alpha^{4/3}\omega \Lambda^{2/3}}
\eeq

\beq %
\ell = \frac{(96)^{1/6}\Lambda^{1/3}}{\pi \alpha^{5/6}\omega^{3/2}}
\eeq

\beq %
\delta D =\frac{1}{\pi}\left(\frac{3}{2}\right)^{4/3}\alpha^{1/3}\Lambda^{2/3}\ln\left(\frac{L}{\ell}\right)
\eeq

\beq %
\xi =\sqrt{\text{exp}\left(\frac{4\left(2/3\right)^{5/3}}{\alpha ^{5/3} \Lambda^{4/3} \omega } \right)-1}
\eeq

\beq %
\nu =\frac{(2/3)^{4/3}}{2\pi \alpha^{1/3}\Lambda^{2/3}}
\eeq

\subsection{$d_H=1$}
\beq
p_F = \frac{2^{1/4}\sqrt{\omega }}{\left(\sqrt{12 \alpha  \ln \left(\frac{\Lambda }{\sqrt{\omega }}\right)+3 \alpha +1}+1\right)^{1/4}}
\eeq

\beq %
D_c=\frac{48 p_F^2 \left(3 \alpha +24 \alpha  \ln \left(\frac{\Lambda }{p_F}\right)+2 \sqrt{3 \alpha +12 \alpha  \ln \left(\frac{\Lambda }{p_F}\right)+1}+2\right)}{\alpha ^2 \omega  \left(33 \alpha  \left(8 \ln \left(\frac{\Lambda }{p_F}\right)+1\right)+304 \pi  \sqrt{3 \alpha +12 \alpha  \ln \left(\frac{\Lambda }{p_F}\right)+1}+22 \sqrt{3 \alpha +12 \alpha  \ln \left(\frac{\Lambda }{p_F}\right)+1}+22\right)}
\eeq
\beq%
\ell = \frac{3 \left(3 \alpha +24 \alpha  \ln \left(\frac{\Lambda }{p_F}\right)+2 \sqrt{3 \alpha +12 \alpha  \ln \left(\frac{\Lambda }{p_F}\right)+1}+2\right)}{19 \pi  \alpha ^2 p_F \sqrt{3 \alpha +12 \alpha  \ln \left(\frac{\Lambda }{p_F}\right)+1}}
\eeq

\beq %
\delta D =\frac{152 p_F^2 \left(3 \alpha +24 \alpha  \ln \left(\frac{\Lambda }{p_F}\right)+2 \sqrt{3 \alpha +12 \alpha  \ln \left(\frac{\Lambda }{p_F}\right)+1}+2\right)}{\omega  \left(33 \alpha  \left(8 \ln \left(\frac{\Lambda }{p_F}\right)+1\right)+304 \pi  \sqrt{3 \alpha +12 \alpha  \ln \left(\frac{\Lambda }{p_F}\right)+1}+22 \sqrt{3 \alpha +12 \alpha  \ln \left(\frac{\Lambda }{p_F}\right)+1}+22\right)} \ln\left(\frac{L}{\ell}\right)
\eeq

\beq %
\nu= \frac{\omega  \sqrt{3 \alpha +12 \alpha  \ln \left(\frac{\Lambda }{p_F}\right)+1}}{\pi  p_F^2 \left(3 \alpha +24 \alpha  \ln \left(\frac{\Lambda }{p_F}\right)+2 \sqrt{3 \alpha +12 \alpha  \ln \left(\frac{\Lambda }{p_F}\right)+1}+2\right)}
\eeq

\beq %
\xi = \sqrt{e^{\frac{12}{19 \alpha ^2}}-1}
\eeq

\subsection{$d_H=2$}
\beq
p_F(\alpha \ll \omega) = \sqrt{\frac{1}{2} \left(\sqrt{9 \alpha ^2+4 \omega ^2}-3 \alpha \right)}
\eeq
\beq
p_F(\alpha \gg \omega) = \frac{ \sqrt{2\omega }}{(6 \ln \left(\frac{\alpha }{\omega }\right)+7)^{1/4}}
\eeq

\beq %
\label{eq: dc dh2 a}
D_c(\alpha \ll \omega)=
\frac{2 \sqrt{9 \alpha ^2+4 \omega ^2} \left(\sqrt{9 \alpha ^2+4 \omega ^2}-3 \alpha \right)^3}{3 \alpha ^2 \omega  \left(\sqrt{9 \alpha ^2+4 \omega ^2}+4 \pi  \left(\sqrt{9 \alpha ^2+4 \omega ^2}-3 \alpha \right)\right)},
\eeq

\beq %
\label{eq: dc dh2 b}
D_c(\alpha \gg \omega) =\frac{32 \omega ^2 \left(3 \ln \left(\frac{\alpha  \sqrt{6 \ln \left(\frac{\alpha }{\omega }\right)+7}}{2 \omega }\right)+2\right) \left(6 \ln \left(\frac{\alpha  \sqrt{6 \ln \left(\frac{\alpha }{\omega }\right)+7}}{2 \omega }\right)+5\right)^2}{3 \alpha ^2 \left(6 \ln \left(\frac{\alpha }{\omega }\right)+7\right)^{3/2} \left(3 \ln \left(\frac{\alpha  \sqrt{6 \ln \left(\frac{\alpha }{\omega }\right)+7}}{2 \omega }\right)+8 \pi +2\right)} \sim \frac{\omega^2}{\alpha^2},
\eeq

\beq %
\ell(\alpha \ll \omega)= \frac{\sqrt{18 \alpha ^2+8 \omega ^2} \sqrt{\sqrt{9 \alpha ^2+4 \omega ^2}-3 \alpha }}{3 \pi  \alpha ^2} \sim \frac{\omega^{3/2}}{\alpha^2}
\eeq

\beq %
\ell(\alpha \gg \omega) = \frac{2 \sqrt{2} \omega ^{3/2} \left(3 \ln \left(\frac{\alpha  \sqrt{6 \ln \left(\frac{\alpha }{\omega }\right)+7}}{2 \omega }\right)+2\right) \left(6 \ln \left(\frac{\alpha  \sqrt{6 \ln \left(\frac{\alpha }{\omega }\right)+7}}{2 \omega }\right)+5\right)}{3 \pi  \alpha ^2 \left(6 \ln \left(\frac{\alpha }{\omega }\right)+7\right)^{3/4}} \sim \frac{\omega^{3/2}}{\alpha^2}
\eeq

\beq %
\delta D(\alpha \ll \omega) =\frac{16 \omega  \sqrt{9 \alpha ^2+4 \omega ^2}}{3 \alpha  \left(\sqrt{9 \alpha ^2+4 \omega ^2}+3 \alpha \right)+4 (1+4 \pi ) \omega ^2} \ln\left(\frac{L}{\ell}\right)
\eeq

\beq %
\delta D(\alpha \gg \omega) =\frac{16 \left(3 \ln \left(\frac{\alpha  \sqrt{6 \ln \left(\frac{\alpha }{\omega }\right)+7}}{2 \omega }\right)+2\right)}{\sqrt{6 \ln \left(\frac{\alpha }{\omega }\right)+7} \left(3 \ln \left(\frac{\alpha  \sqrt{6 \ln \left(\frac{\alpha }{\omega }\right)+7}}{2 \omega }\right)+8 \pi +2\right)} \ln\left(\frac{L}{\ell}\right)
\eeq

\beq %
\xi(\alpha \ll \omega) =\ell\sqrt{\text{exp}\left(\frac{\left(\sqrt{9 \alpha ^2+4 \omega ^2}-3 \alpha \right)^2}{3 \alpha ^2}\right)-1}
\eeq

\beq %
\xi(\alpha \gg \omega) = \ell
\sqrt{\exp \left(\frac{4 \omega ^2 \left(6 \ln \left(\frac{\alpha  \sqrt{6 \ln \left(\frac{\alpha }{\omega }\right)+7}}{2 \omega }\right)+5\right)^2}{3 \alpha ^2 \left(6 \ln \left(\frac{\alpha }{\omega }\right)+7\right)}\right)-1}
\eeq

\beq %
\nu(\alpha \ll \omega) = \frac{\omega }{2 \pi  \sqrt{9 \alpha ^2+4 \omega ^2}}
\eeq

\beq %
\nu(\alpha \gg \omega) =\frac{\sqrt{6 \ln \left(\frac{\alpha }{\omega }\right)+7}}{4 \pi  \left(3 \ln \left(\frac{\alpha  \sqrt{6 \ln \left(\frac{\alpha }{\omega }\right)+7}}{2 \omega }\right)+2\right)}
\eeq

\end{widetext}

\bibliography{referencesMembrane}
\end{document}